\documentclass{pasj01}
\usepackage{lscape}
\usepackage{booktabs}

\begin{document} 
\Received{2015/12/1}
\Accepted{2016/4/2}

\title{Optical Variability Properties of Mini-BAL and NAL Quasars}

\author{Takashi \textsc{Horiuchi}\altaffilmark{1}}
\altaffiltext{1}{Department of Physics, Faculty of Science, Shinshu
  University, 3-1-1 Asahi, Matsumoto, Nagano 390-8621}
\email{th.uchu.im@gmai.com}

\author{Toru \textsc{Misawa}\altaffilmark{2}} \altaffiltext{2}{School
  of General Education, Shinshu University, 3-1-1 Asahi, Matsumoto,
  Nagano 390-8621}

\author{Tomoki \textsc{Morokuma}\altaffilmark{3}}
\altaffiltext{3}{Institute of Astronomy, Graduate School of Science,
  The University of Tokyo, 2-21-1 Osawa, Mitaka, Tokyo 181-0015}

\author{Suzuka \textsc{Koyamada}\altaffilmark{1}}

\author{Kazuma \textsc{Takahashi}\altaffilmark{1}}

\author{Hisashi \textsc{Wada}\altaffilmark{1}}



\maketitle

\begin{abstract}
While narrow absorption lines (NALs) are relatively stable, broad
absorption lines (BALs) and mini-BAL systems usually show violent time
variability within a few years via a mechanism that is not yet
understood. In this study, we examine variable ionization state (VIS)
scenario as a plausible mechanism, as previously suspected. Over three
years, we performed photometric monitoring observations of four
mini-BAL and five NAL quasars at $z_{\rm em}$ $\sim$ 2.0 -- 3.1 using
the 105 cm Kiso Schmidt Telescope in $u$, $g$, and $i$-bands. We also
performed spectroscopic monitoring observation of one of our mini-BAL
quasar (HS1603+3820) using the 188-cm Okayama Telescope over the same
period as the photometric observations. Our main results are as
follows: (1) Structure function (SF) analysis revealed that the quasar
UV flux variability over three years was not large enough to support
the VIS scenario, unless the ionization condition of outflow gas is
very low.  (2) There was no crucial difference between the SFs of
mini-BAL and NAL quasars. (3) The variability of the mini-BAL and
quasar light curves was weakly synchronized with a small time delay
for HS1603+3820. These results suggest that the VIS scenario may need
additional mechanisms such as a variable shielding by X-ray warm
absorbers.
\end{abstract}

\section{Introduction}

Quasars are useful background sources when investigating objects along
our lines of sight. The absorption features in quasar spectra (i.e.,
quasar absorption lines; QALs) are usually classified into {\it
  intervening} QALs, which originate in intervening galaxies and the
intergalactic medium, and {\it intrinsic} QALs, whose origin is
physically associated to the background quasars. The latter comprise
the accelerated gas outflow from the quasars themselves.

The gas outflow can be accelerated by several possible mechanisms:
radiation pressure in the lines and continuum
\citep{1995ApJ...451..498M,2000ApJ...543..686P}, magnetocentrifugal
force \citep{2005ApJ...631..689E}, and thermal pressure
\citep{2005ApJ...625...95C}.  However, the primary mechanism of the
gas outflow is poorly understood. The outflow winds are important
because (1) they eject angular momentum from the quasar accretion disk
and promote accretion of new gas
\citep{1995ApJ...451..498M,2000ApJ...543..686P}, (2) they expel large
amounts of energy and metallicity, thus contributing to the chemical
evolution of the local universe
\citep{2007A&A...463..513M,2005Natur.433..604D}, and (3) they regulate
star formation in nearby interstellar and intergalactic regions.

Broad absorption lines (BALs), defined as lines with a full width at
half maximum (FWHM) exceeding 2,000 ~km s$^{-1}$
\citep{1991ApJ...373...23W}, have been routinely used in outflow wind
studies. However, the line parameters (e.g., column density and line
width) of BALs cannot be measured by model fitting because the line
profiles are hopelessly blended and saturated. On the other hand,
mini-BALs (with FWHMs of 500 --- 2,000 km s$^{-1}$) and narrow
absorption lines (NALs; with FWHMs $\leq$ 500~km s$^{-1}$) contain
internal structures that can be model-fitted to probe their properties
(e.g. \cite{2005ApJ...629..115M}). The observed BALs, mini-BALs, or
NALs depend on the viewing angle to the outflow stream
\citep{1995ApJ...451..498M,2001ApJ...549..133G}. The detection rates
of BALs, mini-BALs, and NALs are $\sim$10-15\%, $\sim$5\%, and
$\sim$50\%, respectively (\cite{2012ASPC..460...47H}), which probably
indicate the global covering fraction of the absorbers around the
continuum sources.

Around 70 -- 90\%\ of BALs are time-variable within 10 years
\citep{2008ApJ...675..985G,2011MNRAS.413..908C}. As an extreme case,
the measured C$\emissiontype{IV}$ BAL variability of
SDSSJ141007.74+541203.3 is only 1.20 days in the quasar rest-frame
\citep{2015ApJ...806..111G} representing the shortest timescale of
absorption line variability ever reported. Recently,
\citet{2014ApJ...792...77M} monitored the spectra of mini-BAL and NAL
quasars, and found that only the former shows significant time
variability in its absorption lines.

However, the physical mechanisms of the absorption line variability
remain unclear.  To date, three scenarios have been proposed: (1) gas
clouds crossing our line of sight (the gas motion scenario), (2)
variable attenuation by flux that is redirected toward our line of
sight by scattering material around the quasar (the reflection
scenario), and (3) changing ionization levels in the outflow gas (the
variable ionization state (VIS) scenario).

\citet{2005ApJ...629..115M} spectroscopically monitored the
C$\emissiontype{IV}$ mini-BAL in the quasar HS~1603+3820 for more than
four years. They found multiple troughs in the mini-BAL that vary in
concert. This finding eliminates the gas motion scenario (at least in
1603+3820) because it implies simultaneous crossing of gas clouds over
our line of sight, which is unlikely. \citet{2010ApJ...719.1890M} also
rejected the reflection scenario, because in spectropolarimetric
observations of the same mini-BAL system, the fraction of polarized
flux (i.e., the flux redirected by scattering material) is only
$\sim$0.6~\%, too small to support the reflection scenario.
\citet{2008ApJ...675..985G} found no correlations between quasars and
absorption lines variability in 13 BAL quasars. On the other hand,
\citet{2013A&A...557A..91T} simultaneously monitored the equivalent
widths (EWs) of BALs and the ultraviolet (UV) luminosities of their
host quasars (i.e., ionizing photon density) and found clear
correlations in a single quasar, supporting the VIS scenario. The VIS
scenario has not been tested in mini-BAL / NAL quasars and is still
being debated.

In this study, we verify the VIS scenario in the light curves of four
mini-BAL quasars and five NAL quasars (hereafter, quasar
variability\footnote[1]{On the other hand, changes in the absorption
  strength/feature are referred to as the ``absorption line
  variability''}). We also search for possible correlations between
the outflow and quasar parameters, as discussed in the literature
(e.g.  \cite{1999MNRAS.306..637G} (G99, hereafter);
\cite{2004ApJ...601..692V} (VB04, hereafter);
\cite{2005AJ....129..615D}; \cite{2007MNRAS.375..989W};
\cite{2008MNRAS.383.1232W} (W08, hereafter);
\cite{2013A&A...560A.104M}). Section 2 of this paper describes the
sample selection, observation, and data analysis. In Section 3, we
present the photometric data of mini-BAL / NAL quasars.  Section 4
discusses the viability of the VIS scenario in mini-BAL and NAL
quasars and the possible correlations between parameters. Results are
summarized in Section 5. Throughout, we adopt a cosmological model
with $H_{0}$=70~km s$^{-1}$~Mpc$^{-1}$, $\Omega_{m}$=0.27 and
$\Omega_{\Lambda}$=0.73.

\section{Observation and Data Analysis}

\subsection{Sample Selection}

 Our samples are selected based on availability of multi-epoch high
 dispersion spectroscopic studies in \citet{2014ApJ...792...77M}. 
 We sampled four mini-BAL quasars (HS1603+3820, Q1157+014,
  Q2343+125, and UM675) and five NAL quasars (Q0450-1310\footnote[2]{
  Although this quasar was not studied in \citet{2014ApJ...792...77M}, 
  we sampled it because it hosts a reliable {\it intrinsic} NAL confirmed by 
  \citep{2007ApJS..171....1M}.} Q0940-1050, Q1009+2956, Q1700+6416, 
  and Q1946+7658), whose absorption line variabilities (or non-variabilities) have 
  been already studied by \citet{2014ApJ...792...77M} using Subaru with
  the High Dispersion Spectrograph (HDS, $R \sim$ 45,000), Keck with 
  the High Resolution Echelle Spectrometer (HIRES, $R \sim$ 36,000), 
  and Very Large Telescope (VLT) with the Ultraviolet and Visual Echelle 
  Spectrograph (UVES, $R \sim$ 40,000) in time intervals of $\sim$ 4 $-$ 
  12 years. Our sample quasars are summarized in Table~1.

\subsection{Imaging Observations}

Photometric observations were performed by the 105-cm Kiso Schmidt
Telescope with a Kiso Wide Field Camera (KWFC, \cite{S12}).  The eight
2K$\times$4K charge coupled devices (CCDs) in the KWFC provides a
field-of-view (FoV) of 2.2$^{\circ}$ $\times$ 2.2$^{\circ}$. Since
five of our nine quasars are located in the Sloan digital sky survey
(SDSS) field, our photometry used the SDSS ($u$, $g$, and $i$) filters
instead of the Johnson filters. Moreover, as the $u$-band is less
sensitive than the $g$- and $i$-bands, we adopted a 2 $\times$ 2
binning mode (1.89 arcsec/pixel) for the $u$-band observations.

The quasars were repeatedly observed from April 14, 2012 to October
16, 2014, with a typical monitoring interval of three months,
representing the typical variability time scale of BALs (e.g.,
\cite{2011MNRAS.413..908C}). Observation logs of the individual
quasars are summarized in Table~2. The log excludes Q0450-1310 and
Q1946+7658 in the $u$-band because the continuum fluxes of these
quasars are heavily absorbed by the foreground intergalactic medium
(i.e., Ly$\alpha$ forest). Bias subtraction, flat-fielding, sky
subtraction, and World coordinate system matching were performed by an
automatic analysis pipeline. The same pipeline was used for supernova
discoveries in the Kiso Supernova Survey (KISS) project
\citep{2014PASJ...66..114M}.

\subsection{Relative Photometry}

The extraction and magnitude measurements of quasars and comparison
stars were performed by SE{\footnotesize XTRACTOR}
\citep{1996A&AS..117..393B}. Regions crowded with stars were selected
by the flux estimation code ${\tt FLUX\_BEST}$.

Since we mainly investigate the light curves of quasars (i.e., the
{\it relative} magnitudes between observing epochs), we do not need to
measure their {\it true} magnitudes. Therefore, we performed relative
photometry by simultaneously monitoring the quasars and {\it
  effective} photometric standard stars (hereafter called {\it
  comparison stars}) near the quasars. The comparison stars were
selected as follows. We chose two (unsaturated) bright stars near the
target quasars in the same CCDs and investigated their {\it relative}
magnitudes $\Delta m$ (= $|m_{s1} - m_{s2}|$), where $m_{s1}$ and
$m_{s2}$ are the magnitudes of the bright stars. If their 
  relative variability between the two stars $| \Delta m$ $-$
$\langle\Delta m \rangle |$, where $\langle\Delta m \rangle$ is
  the average value of all observations, was always below 0.05 mag
and below the 3$\sigma$ level of the photometric errors (i.e., $\Delta
m$ was very stable), one of the stars was designated a comparison
star. Otherwise, we continued searching for stars that satisfied the
above criteria.  A single comparison star was used in all epochs,
unless different stars in different filters were required.

The quasars were subjected to relative photometry against these
comparison stars and were classified as variable stars if their
magnitude changed by more than 3$\sigma$ and 0.05 mag. The total
photometric error $\sigma_{\rm qso}$ in the quasar photometry (in
units of magnitude) is defined as

\begin{equation}
{\sigma_{\rm qso}}^2 = {\sigma_{\rm ph}}^2 + {\sigma_{\rm star}}^2,
\end{equation}

where ${\sigma_{\rm ph}}$ is the photometric error in the epochs to be
compared and ${\sigma_{\rm star}}$ is the weighted average of the
variability of the comparison star, which is defined as

\begin{equation}
\sigma_{\rm star} = \frac{\sum_{i,i<j} |\Delta m_i - \Delta m_j|
  w_{ij}}{\sum_{i,i<j} w_{ij}},~ w_{ij} = 1/{\sigma_{ij}}^2 .
\end{equation}

In Eq. (2), ${\sigma_{ij}}^2$ is the sum of squares of the photometric
error in the comparison star between epochs $i$ and $j$.

\subsection{Properties of Sample Quasars}

Table~1 lists the properties of our targets, namely, the quasar
parameters (coordinates, emission and absorption redshifts, optical
magnitudes, radio-loudness, bolometric luminosities, black hole
masses, and Eddington ratios) and the absorption parameters (ejection
velocities, whether lines are variable or not, averaged EWs, and
variability amplitude of EWs).  The last two parameters are measured
for C$\emissiontype{IV}$ absorption lines.  These data were collected
from literature or calculated from the reported data. After
calculating the monochromatic luminosity at $\lambda$ = 1450\AA\ from
the V-band magnitude, we applied the bolometric correction $L_{bol}$
$\sim$ 4.4$\lambda L_{\lambda}$, following
\citet{2004ApJ...601..715N}. For the black hole mass, we used the
heuristic equation of \citet{2006ApJ...641..689V},

\begin{equation}
{\rm log}\left(\frac{M_{\rm BH}}{M_{\solar}}\right)=0.660+0.53{\rm log}
\left(\frac{\lambda L_{\lambda}}{10^{44}~\rm erg/s}\right)+2{\rm log}
\left(\frac{\rm FWHM}{\rm km~s^{-1}}\right),
\end{equation}

where the FWHM of the C$\emissiontype{IV}$ broad emission line is
measured from VLT/UVES archive spectra.

The quasar parameters of our targets were compared with those of
$\sim$17,000 quasars at $z_{\rm em}$ $\sim$ 2.0 -- 3.1 from the SDSS
Data Release 7 (SDSS DR7) (see Figure~1). Our quasars demonstrate
extremely large luminosity with a mean $\langle L_{bol} \rangle =2.29
\times 10^{48}$ ergs~s$^{-1}$. Eight of our quasars qualify as super
Eddington with a mean Eddington ratio of $\langle \varepsilon \rangle
= 3.02$, although their black hole masses are comparable to those of
the SDSS quasars in the same redshift range. The mean quasar
luminosity and Eddington ratio of SDSS DR7 (cataloged by
\cite{2011ApJS..194...45S}) are 5.13$\times 10^{46}$ erg~s$^{-1}$ and
0.41 respectively.

The radio-loudness $R=f_{\nu}(5{\rm GHz})/f_{\nu}(4400{\rm \AA})$ was
also collected from the literature or calculated from FIRST radio
measurements.  Two quasars (Q1157+014 and UM675) are classifiable as
radio-loud ($R~>~10$; \cite{1989AJ.....98.1195K}), while the other 7
quasars are radio-quiet.

\subsection{Spectroscopic Observation for HS1603+3820}
We also performed spectroscopic monitoring observations of a single
mini-BAL quasar (HS1603+3820) using the 188-cm Okayama Telescope with
a Kyoto Okayama Optical Low-dispersion Spectrograph (KOOLS;
\cite{2005JKAS...38..117Y}).  For these observations, we selected a
VPH495 prism, which is sensitive to 4,500-5,400\AA~and a
$\timeform{1.^{!!\prime\prime}8}$ slit (yielding $R \sim $1,100). The
CCD was binned every 2 $\times$ 2 pixels.

Observations were performed from September 19, 2012 to May 21, 2015
over typical monitoring intervals of three months. Useful data were
acquired on September 19 of 2012, May 30 of 2015, February 23 of 2015,
and May 21 of 2015 (hereafter, these four periods are referred to as
epochs 1, 2, 3, and 4). The observing log is listed in Table~3.

\section{Results}

This section present the photometric variability results of each
quasar determined from light curves. The quasar variability properties
of the mini-BAL and NAL quasars are then compared by SFs and color
variability analysis. The results are summarized in Figures~2 and 3
and in Table~4.

\subsection{Quasar variability}
To examine the quasar variability of the nine mini-BAL / NAL quasars,
we measured the standard deviation in the magnitude $\sigma_m$, the
mean quasar variability $\langle|\Delta m|\rangle$, the maximum
magnitude variability $|\Delta m|_{\rm max}$, the mean quasar
variability gradient $\langle|\Delta m/\Delta t_{\rm rest}|\rangle$,
and the maximum quasar variability gradient $|\Delta m/\Delta t_{\rm
  rest}|_{\rm max}$, following \citet{1994A&A...284..764B} and
G99. The mean values were calculated from all combinations of the
observing epochs (e.g., from $_NC_2$ combinations , where $N$ is the
number of observing epochs.). The quasar variability gradient was
defined as the quasar variability per unit time (year). These
parameters are summarized in Table~4. The maximum quasar variability
and its gradient are listed even if their significance level is below
3$\sigma$.

The most remarkable trend is the larger quasar variabilities in bluer
bands than those in redder bands. This well-known property of quasars
is repeatedly discussed in literature
(e.g.,\cite{1997A&A...321..123C}; VB04; \cite{2012ApJ...758..104Z};
\cite{2014ApJ...792...33G}).  The largest quasar variabilities were
exhibited by HS1603+3820 among the mini-BAL quasars ($|\Delta u_{\rm
  max}|$ $\sim$0.23) and by Q1700+6416 among the NAL quasars ($|\Delta
u_{\rm max}|$ $\sim$0.30), while the largest variability gradients
were exhibited by Q1157+014 among the mini-BAL quasars ($|\Delta
i/\Delta t_{\rm rest}|_{\rm max}$ $\sim$5.0) and by Q1946+7658 among
the NAL quasars ($|\Delta g/\Delta t_{\rm rest}|_{\rm max}$
$\sim$16.9).

\subsection{Notes on Individual Quasars}

HS1603+3820 (mini-BAL, $z_{em}$=2.542, $m_{\rm V}$=15.9) --- This
quasar exhibited a violently variable mini-BAL profile with an
ejection velocity $v$ $\sim$ 9,500~km s$^{-1}$
\citep{2007ApJ...660..152M}. Among the mini-BAL quasars in the present
study, this quasar showed the largest variability in the $u$-band
($|\Delta u|$ $\sim$ 0.23~mag) and the second largest variability in
the $g$-band ($|\Delta g|$ $\sim$ 0.19~mag) among our mini-BAL
quasars.  On the other hand, the mean and maximum quasar variability
of HS1603+3820 were surprisingly small in the $i$-band (only
$\sim$0.01 and $\sim$0.05~mag, respectively).
For this quasar alone, we supplemented the photometric observations
with spectroscopic observations. Obtained C$\emissiontype{IV}$
mini-BALs in this quasar in each epochs are summarized in Figure~4,
and we measured the EW of the C$\emissiontype{IV}$ mini-BAL and
monitored its variability.  The results are summarized in Figure~5 and
Table~5. The EW marginally varied between epochs 1 and 3 with
absorption variability amplitude $\Delta{\rm EW}$ = 6.0 $\pm$4.2${\rm
  \AA}$ (significance level $\sim$1.5$\sigma$).

Q1157+014 (mini-BAL, $z_{em}$=2.00, $m_{\rm V}$=17.6) --- This
radio-loud quasar ($R=471$) was the faintest among our sample
quasars. At the start of our monitoring campaign, Q1157+014 showed a
rapid quasar variability in the $i$-band with an amplitude $|\Delta
i|$ $\sim$ 0.14~mag, much larger than those of the $u$- and $g$-band,
between the first (April 2012) and second (May 2012)
epochs. Thereafter, the magnitude variability remained high in the
$u$-band and reduced in the $i$-band.

Q2343+125 (mini-BAL, $z_{em}$=2.515, $m_{\rm V}$=17.0) --- This quasar
exhibited the largest Eddington ratio $\varepsilon$ among our mini-BAL
quasars ($\varepsilon$ $\sim$4.90) and the smallest mean quasar
variability in the $g$-band ($\langle |\Delta g| \rangle \sim$
0.02). The quasar variability was only slightly larger in the $i$-band
than in the $g$-band. Although Q2343+125 was observed only twice in
the $u$-band, precluding an evaluation of its variability trend in
that band, it appears that the quasar variability trends were
consistent in all three bands.

UM675 (mini-BAL, $z_{em}$=2.15, $m_{\rm V}$=17.1) --- This radio-loud
quasar ($R=438$) has a sub-Eddington luminosity ($\varepsilon=$0.91)
and exhibited the largest variability in the $g$- and $i$-band among
the mini-BAL quasars ($|\Delta g|$ and $|\Delta i|$ are $\sim$0.22 and
0.16~mag, respectively).  Similar to Q2343+125, detailed trends in the
$u$-band were precluded by the limited number of monitoring epochs.

Q0450-1310 (NAL, $z_{em}$=2.30, $m_{\rm V}$=16.5) --- The magnitude of
this quasar suddenly changed ($|\Delta g|$ $\sim$0.16~mag) in the
$g$-band during the last three months of observations (from September
2013 to December 2013).  The $\Delta m$ in the $g$- and $i$-band
largely differed from the 3rd to the 5th observing epochs, possibly
because there were few observing epochs in the $i$-band.

Q0940-1050 (NAL, $z_{em}$=3.080, $m_{\rm V}$=16.6) --- The $g$- and
$i$-band fluxes monotonically decreased during the monitoring
campaign. The quasar variability amplitudes of the all bands were
almost identical. In this case, the variable trend in the $u$-band was
obscured by the large photometric error, especially in the 2nd
epoch. These errors were introduced by bad weather.

Q1009+2956 (NAL, $z_{em}$=2.644, $m_{\rm V}$=16.0) --- Among our
samples, this NAL quasar has the largest Eddington ratio
($\varepsilon=$7.21) and the smallest variability level in all bands
($|\Delta m|$ $\leq$ 0.06~mag).

Q1700+6416 (NAL, $z_{em}$=2.722, $m_{\rm V}$=16.13) --- The bolometric
luminosity and black hole mass of this quasar were the largest among
our samples. Q1700+6416 also exhibited the largest $u$-band
variability ($|\Delta u|$ $\sim$0.3~mag) among our samples.

Q1946+7658 (NAL, $z_{em}$=3.051, $m_{\rm V}$=15.85) --- This quasar
exhibited a cyclic quasar variability pattern with the highest
half-year variability of the $g$-band magnitude in the quasar
rest-frame ($|\Delta g|$ $\sim$0.24~mag). Conversely, the $i$-band
magnitude was very stable over the same observation term.

\subsection{Structure Function Analysis}
We now examine the effects of time-scale and wavelength on the quasar
variability properties. These relationships are usually determined
through SF analysis. For this purpose, we adopt the Structure Function
(SF) proposed by \citet{1996ApJ...463..466D},

\begin{equation}
S = \displaystyle\sqrt{\frac{\pi}{2} \left\langle|\Delta m (\Delta
  \tau)|\right\rangle^2-\left\langle \sigma_{n}^2 \right\rangle},
\end{equation}

where $|\Delta m (\Delta \tau)|$ and $\sigma_{n} =
\sqrt{\sigma_{m_i}^2+ \sigma_{m_j}^2}$ are the magnitude variability
and its photometric error, respectively, between two observing epochs
separated by $\Delta \tau = t_j - t_i$ in the quasar's rest-frame. The
bracket denotes the averaged value in paired observing epochs with
time-lags within a specific range (after separation into four
bins). Note that the wavelength coverage of the $u$-, $g$-, and
$i$-band in the rest frame depends on the quasar redshift (see
Figure~6).  However, because the distribution of the emission redshift
was $\sim$2.5 in two-thirds of our samples (six out of nine quasars),
we can investigate the wavelength dependence of the SF. Detailed
trends are investigated later in this subsection.

Figure~7 plots the SF as a function of time lag in the rest-frame for
the $u$-, $g$-, and $i$-band. In all bands, the quasar variability
increases with the time lag $\Delta \tau$. Because the observing
epochs were fewer in the $u$- and $i$-band than in the $g$-band, they
introduce larger errors in the SF. The SF is often fitted to a
power-law (\cite{1994MNRAS.268..305H}; \cite{2002ApJS..141...45E};
VB04; W08):
 
\begin{equation}
S_p(\Delta \tau)=\displaystyle \left( \frac{\Delta \tau}{\Delta
  \tau_p} \right)^\gamma,
\end{equation}

where $\gamma$ is the power-law index and the time scale $\Delta
\tau_p$ defined such that $S_p(\Delta \tau_p)$ equals 1 mag. The
fitting parameters to this model are summarized in Table~6. Note that
because $\Delta \tau_p$ has an extremely large uncertainty, especially
for small samples such as ours (see W08), we replace it by
$S_p$($\Delta \tau =100$~days). The SFs were also fitted to the
following asymptotic function (e.g., \cite{1994ApJ...433..494T};
\cite{1994MNRAS.268..305H}; \cite{2002ApJS..141...45E}):

\begin{equation}
S_a(\Delta \tau)=\displaystyle V_a(1-e^{-\Delta \tau/\Delta \tau_a}),
\end{equation}

where $V_a$ is the asymptotic value at $\Delta \tau = \infty$. Table~6
lists the best-fit parameters to this function, along with those of
W08 and VB04. \footnote[3]{We emphasize that the rest-frame wavelength
  regions studied in the current work may differ from those in the
  literature.} In all cases, the quasar variability is higher at bluer
than at redder wavelengths.

Finally, to examine the wavelength dependence of SF, we fitted the SF
to the following equation (VB04):

\begin{equation}
S(\lambda)=\displaystyle A {\exp}(-\lambda/\lambda_0)+B,
\end{equation}

where $A, B$ and $\lambda_0$ are fit parameters. First, we separated
our mini-BAL and NAL samples using a boundary time-lag of $\Delta
\tau$ = 90~days\footnote[4]{The average time lag of all combinations
  of observing epochs in all bands (used as the criterion).}  in the
rest-frame, then fitted the subsamples to the above model. The fitting
curves of our data and VB04's data are plotted in Figure~8. The quasar
variability clearly decreases with wavelength, as noted in literature
(e.g., G99; VB04; \cite{2005AJ....129..615D};
\cite{2012ApJ...758..104Z}). Moreover, the magnitudes of our SF are
much lower than those of VB04's data because our mini-BAL / NAL
quasars were much brighter than normal SDSS quasars in the same
redshift range (Figure~1). The trend of the fitting reflects the
anti-correlation between quasar variability and luminosity. No clear
differences are observed between mini-BAL and NAL quasars.

\subsection{Color Variability}
Color variability is among the most remarkable properties of
quasars. Although our relative photometry cannot determine the {\it
  true} magnitudes of quasars (see section 2.3), the color variability
can be evaluated through the cancellation of photometry shifts
($\delta m$). For example, we can write

\begin{eqnarray}
 \Delta (u-g) & = & (u_2 +\delta u) - (g_2 + \delta g) - (u_1 +\delta u) +
 (u_1 + \delta g)\\
              & = & u_2 - u_1 - (g_2 - g_1)\\
              & = & \Delta u - \Delta g,
\end{eqnarray}

where the subscripts on $u_1$ and $u_2$ denote the first and second
observing epochs in the comparison.

Figure~9 plots the $\Delta (u-g)$, $\Delta (u-i)$, and $\Delta (g-i)$
color variabilities as functions of quasar variability. The
correlation properties of the mini-BAL and NAL quasars are summarized
in Table~7. The color and magnitude variabilities are positively
correlated in both mini-BAL and NAL quasars (namely, brighter quasars
tend to be bluer; hereafter called the BWB trend). The same phenomenon
has been reported in normal quasars (e.g. G99;
\cite{2000ApJ...540..652W}; VB04; \cite{2010ApJ...711..461S}, 2011;
\cite{2014ApJ...783...46K}). The correlation trends are consistent in
the mini-BAL and NAL quasars.

The standard deviations of the quasar colors, the mean and maximum
color variabilities, and the mean and maximum color variability
gradients of the mini-BAL and NAL quasars, are listed in
Table~8. Again, no significant differences exist between the mini-BAL
and NAL quasars, except for 5.4$\sigma$ difference in the maximum
color gradient of $\Delta (u-i)$.

\section{Discussion}

\subsection{Quasar Variability Trends of Mini-BAL and NAL Quasars}

\subsubsection{Structure Function}
Comparing the SF fitting parameters of the mini-BAL and NAL quasars to
those of normal quasars reported in VB04 and W08 (Table~6), we observe
the following trends:

\begin{itemize}
\item[(i)] The power-law indices $\gamma$ of the mini-BAL and NAL
  quasars ($\gamma$ $\sim ~$0.410$\pm$0.115, 0.264$\pm$0.056, and
  0.436$\pm$0.115) were consistent with those of normal quasars
  reported in W08 ($\gamma$ $\sim$0.43, 0.48, and 0.44) except in the
  $g$-band, although the rest-frame wavelength coverage differed
  among the quasar samples (being dependent on the redshift
  distribution of the quasar). Similar indices were obtained in a disk
  instability model \footnote[5]{Transient flares or blob formations
    caused by any instability should alter the luminosity.} ($\gamma =
  0.41 \sim 0.49$; \cite{1998ApJ...504..671K}). No significant
  differences were observed between the mini-BAL and NAL quasars.

\item[(ii)] In the asymptotic model $V_a$, the asymptotic value at
  $\Delta \tau$ = $\infty$ of mini-BAL / NAL quasars was approximately
  half that of normal quasars in the $g$ and $i$-band. The same
  phenomenon was observed for $S$($\Delta \tau$ = 100~days).

\end{itemize}


\subsubsection{Color variability}
The mini-BAL and NAL quasars exhibit similar color-magnitude
variability (Table~7) and color variability (Table~8) with one
exception: a 5.4$\sigma$ difference in the maximum color gradients
(MCGs; ($|\Delta C / \Delta t_{\rm rest}|$)$_{\rm max}$).

The 5.4$\sigma$ difference in MCGs was observed between a mini-BAL
quasar (HS1603+3820) and a NAL quasar (Q1700+6416) with BWB trends in
$\Delta$($u-i$). In both quasars, the variability was maximum in the
$u$-band and moderate in the $i$-band. However, the $u - i$
variability developed over a shorter time~-~frame in Q1700+6416 than
in HS1603+3820, which might explain the larger color variability
gradient in the former than in the latter.

\subsubsection{Correlation between EW and quasar variability}

 As shown in Figure~5, the variability trends of the magnitude and
  EW of the C$\emissiontype{IV}$ mini-BAL for HS1603+3820 were
  marginally synchronized with the quasar variability leading the EW
  variability. Specifically, the EW first increased from 2012
  September (epoch 1) to 2015 February (epoch 3) with a marginal
  significance level of $\sim 1.5 \sigma$ ($\Delta {\rm
    EW}~=~6.0\pm4.2$) and then decreased from 2015 February to 
  2015 May (epoch 4), while the quasar brightness in the $u$-band first decreased 
  from 2012 September to 2014 May and then increased from 2014 May to  
  2015 May. The time-lag of the marginal synchronizing 
  trend in quasar and absorption line variabilities is about nine months 
  ($\sim$2.6 months in the quasar rest-frame). If we assume the time-delay 
  corresponds to the recombination time from C$\emissiontype{V}$ to
  C$\emissiontype{IV}$, we can place a lower limit on the absorber's
  gas density as $n_e \geq~2.8 \times 10^{4}$ cm$^{-3}$ by the same
  prescription as used in \citet{2004ApJ...601..715N}.

  \citet{2013A&A...557A..91T} reported a similar synchronizing
  trend in a BAL quasar APM 08279+5255, although one of two NALs that
  are detected aside the BAL did not show such a synchronization.
  They suggested this was due to a larger recombination time for the
  NAL absorber with smaller electron density compared to the other
  absorbers. Both of these results are not inconsistent to the VIS
  scenario.

\subsection{The VIS Scenario}
Assuming the VIS scenario, we now estimate the quasar variability that
reproduces the observed absorption line variabilities of BAL and
mini-BAL quasars reported in literature. If the VIS scenario holds,
the absorption strengths will depend on the ionization condition of
the absorber, which is quantified by the ionization parameter $U$

\begin{equation}
U \equiv \displaystyle \frac{1}{4\pi r^2 c n_e} \int_{\nu}^{\infty}
\frac{L_{\nu}}{h\nu}d\nu = \frac{Q}{4\pi r^2 c n_e} =
\frac{n_{\gamma}}{n_e},
\end{equation}
where $Q$ is the number density of hydrogen-ionizing photons emitted
from the continuum source per second, $r$ is the distance between the
absorber and the continuum source, and $n_{\gamma}$ and $n_e$ are the
volume densities of the ionizing photons and electrons, respectively.

Here, we assume the absorption line variability of (mini-)BALs is
attributed to recombination to (or ionization from)
C$^{2+}$ ({\it case A}, hereafter) and adopt the
optimal ionization parameters for C$^{2+}$ and
C$^{3+}$ ($\log$U $\sim$ $-$2.8 and $-$2.0,
respectively) \citep{1997ApJS..109..279H}. Because at least one of our
mini-BAL quasars (HS1603+3820) is unlikely to vary by the gas motion
scenario \citep{2005ApJ...629..115M}, we assume constant gas density
$n_e$. Therefore, the ionizing photon density $n_{\gamma}$ should 
increase/decrease by a factor of $\sim$6.3 to change $\log U$ from/to
$-$2.8 to/from $-$2.0, corresponding to $\Delta m$ $\sim$ 2. For
reference, a typical quasar varies by only $\Delta m$ $\sim$ 0.1 over
several months and maximally varies by $\Delta m$ $\sim$ 0.5 over
several years \citep{2000ApJ...540..652W}. These variabilities are
much smaller than the above-required value\footnote[6]{Even if we
  attribute the absorption line variability to recombination to (or
  ionization from) C$\emissiontype{V}$ (whose optimal ionization
  parameter is $\log$U $\sim$ $-$1.2), the required magnitude
  variability would be almost same.}. 

 However, C$\emissiontype{IV}$ absorbers do not necessarily have 
  an optimal ionization parameter for C$^{3+}$ (i.e., $\log$U $\sim$
  $-$2.0). As the other extreme case, if mini-BAL absorbers have $\log$U
  $\sim$ $-$3.0, their ionization fraction $f$ (i.e., a fraction of
  Carbon in ion state C$^{3+}$) is very sensitive to the ionization
  parameter ($\Delta \log$f/$\Delta \log$U $\sim$ 1.8; {\it case B}, hereafter),
  although it weakly depends on the shape of incident ionizing flux.
  Indeed, the value of $\Delta \log$f/$\Delta \log$U for HS1603+3820, which is the  
  only quasar among our sample for which the magnitude and EW of the
  C$\emissiontype{IV}$ mini-BAL were simultaneously monitored over
  three years, is $\sim$1.1 between epochs 1 and 2 ($\Delta \log$EW 
  $\sim \Delta \log {\rm f} \sim$ 0.1)\footnote[7]{If absorbing clouds are 
  optically thin (i.e., absorption lines are at a linear part of the
   curve-of-growth), $\Delta\log$EW is close to $\Delta\log$f, which
   is applicable for all mini-BALs in our sample except for one in
   Q1157+014.} and $\sim$2.0 between the epochs 1 and 3 ($\Delta \log$EW 
  $\sim$ 0.18), assuming $\Delta m$ $\sim$0.23 (the
  maximum quasar variability during our monitoring observations).
  These values are expected for absorbers with ionization parameters
  of $\log$U $\sim$ $-3$ -- $-2$ (see Figure~2 of
  \cite{1997ApJS..109..279H}).
  If this is the case, an averaged amplitude of absorption variability
  in four C$\emissiontype{IV}$ mini-BALs in our sample
  ($\langle\Delta\log$EW$\rangle$ $\sim$ $\langle\Delta\log$f$\rangle$
  $\sim$ 0.1) can be
 caused by only a small change of the ionizing
  flux, $\log$U $\sim$ 0.06. This value corresponds to $\Delta m$
  $\sim$ 0.14, comparable to a typical variability of our sample
  quasars as well as quasars in the literature
  \citep{2000ApJ...540..652W}. The variability amplitude of
  C$\emissiontype{IV}$ ionizing photons in shorter wavelength
  ($\lambda_{\rm rest}$ $\sim$ 200 ${\rm \AA}$) may be even larger
  because of the anti-correlation between quasar variability and
  wavelength (see Section 3.3).

Thus, the case~B is favorable for explaining the variability trend in HS1603+3820
with the VIS scenario. However, it has one shortcoming; four mini-BAL 
systems in our sample have either strong N$\emissiontype{V}$ absorption lines 
or no remarkable Si$\emissiontype{IV}$ absorption lines, which suggests their 
ionization condition is not as low as $\log$U $\sim$ $-3$ (see Figure~2 of 
\cite{1997ApJS..109..279H}). Therefore, it is less likely that the case~B alone 
causes the absorption variability of mini-BALs in our sample quasars.

\subsection{Additional mechanism to support the VIS Scenario}
The outflow wind variability may be caused by more than one
mechanism. We speculate that the VIS scenario is accompanied by an
additional mechanism, such as variable optical depth between the flux
source and the absorber. One promising candidates is a ${\it
  warm~absorber}$ which has been frequently detected in X-ray
spectroscopy (e.g.,\cite{2002ApJ...567...37G}, 2006;
\cite{2007ApJ...659.1022K}; \cite{2012A&A...542A..30M}). Warm
absorbers were originally proposed to avoid over-ionization of the
outflow winds \citep{1995ApJ...451..498M}. Because warm absorbers are
significantly variable in X-ray monitoring observations (e.g.,
\cite{2007AJ....133.1849C}; \cite{2010ASPC..427..108G}, b), the
ionization condition of the UV absorber in the downstream might also
vary. Indeed, in a photoionization model, \citet{2014NewA...28...70R}
estimated that the C$\emissiontype{IV}$ mini-BAL absorber lies within
$r$ = 0.1~pc of the quasar center. Similarly, the X-ray warm absorber
is estimated to be within 0.1 pc of HS1603+3820.
\citet{2001ApJ...549..133G} argued that NAL and BAL absorbers locate
at high and low latitudes above the accretion disk equator,
respectively. A radiation-MHD simulation by
\citet{2013PASJ...65...88T} also predicts no warm absorbers at very
high latitudes. If this picture is correct, X-ray shielding is
ineffective in the NAL outflow directions. Supporting this idea,
X-rays are not strongly absorbed in NAL quasars
\citep{2008ApJ...677..863M}.
The model of \citet{2009ApJ...693.1929K} supports that NAL absorbers
are the interstellar media of host galaxies, which are swept up by the
outflow wind. In this case, the absorbers should exhibit little
variability because their volume density is very small (corresponding
to a very long recombination time). Moreover, they are very distant
(of the order of kpcs) from the continuum source, therefore they
should be weakly influenced by the variable flux source.
However, \citet{2013MNRAS.435..133H} find no evidence of strong X-ray
absorption toward the outflows of either NAL or mini-BAL
quasars. Instead of an X-ray warm absorber, they argue that small
dense clumpy absorbers avoid over-ionization by self-shielding. In
this case, we should expect no correlations between the absorption
strengths of the UV and X-ray fluxes.

\section{Summary}
We performed i) photometric monitoring observations of four mini-BAL
and five NAL quasars over more than three years and ii) spectroscopic
observation for a single mini-BAL quasar (HS1603+3820) to investigate
whether the VIS scenario can explain the absorption line variability
in BALs and mini-BALs. Our main results are summarized below:

\begin{itemize}
\item[(1)] Quasar variability increases with monitoring time-lag but
  decreases with observed wavelength, as previously reported in normal
  quasars.

\item[(2)] Mini-BAL and NAL quasars become bluer as they brightened
  (the BWB trend), as often observed in normal quasars.

\item[(3)] The quasar variability properties did not significantly
  differ between mini-BAL and NAL quasars, indicating that flux and
  color variabilities alone cannot account for the absorption line
  variabilities.

\item[(4)] Quasar magnitude was marginally synchronized with
  absorption strengths in one mini-BAL quasar HS1603+3820, with the
  former temporally leading the latter.

\item[(5)] The VIS scenario cannot causes the absorption variability
  of mini-BALs in our sample quasars unless the ionization condition
  of outflow gas is as low as $\log$U $\sim$ $-$3.

\item[(6)] The VIS scenario may require an additional mechanism that
  regulates incident flux to the outflow gas. The most promising
  candidate is X-ray warm absorbers with variable optical depth.
\end{itemize}

Before conclusively validating the VIS scenario, we need to
simultaneously monitor the outflow and shielding material by UV and
X-ray spectroscopies. The presented monitoring observations should
also be performed on quasars with a wide range of luminosities and
Eddington ratios to mask the anti-correlation effect between the
luminosity/Eddington ratio and quasar variability.

\begin{ack}
We thank Ken'ichi Tarusawa, Takao Soyano and Tsutomu Aoki for
supporting our observations at Kiso Observatory for over three
years. Ikuru Iwata and Hironori Tsutsui support our observations with
the 188-cm Okayama Telescope with KOOLS. We also thank Noboru Ebizuka,
Masami Kawabata and Takashi Teranishi for producing VPH grisms used in
KOOLS and kindly providing us, and Rina Okamoto for supporting our
observations at Kiso Observatory and Okayama Astrophysical
Observatory. The research was supported by JGC-S Scholarship
Foundation and the Japan Society for the Promotion of Science through
Grant-in-Aid for Scientific Research 15K05020.
\end{ack}



\begin{figure*}
 \begin{center}
  \includegraphics[width=16cm]{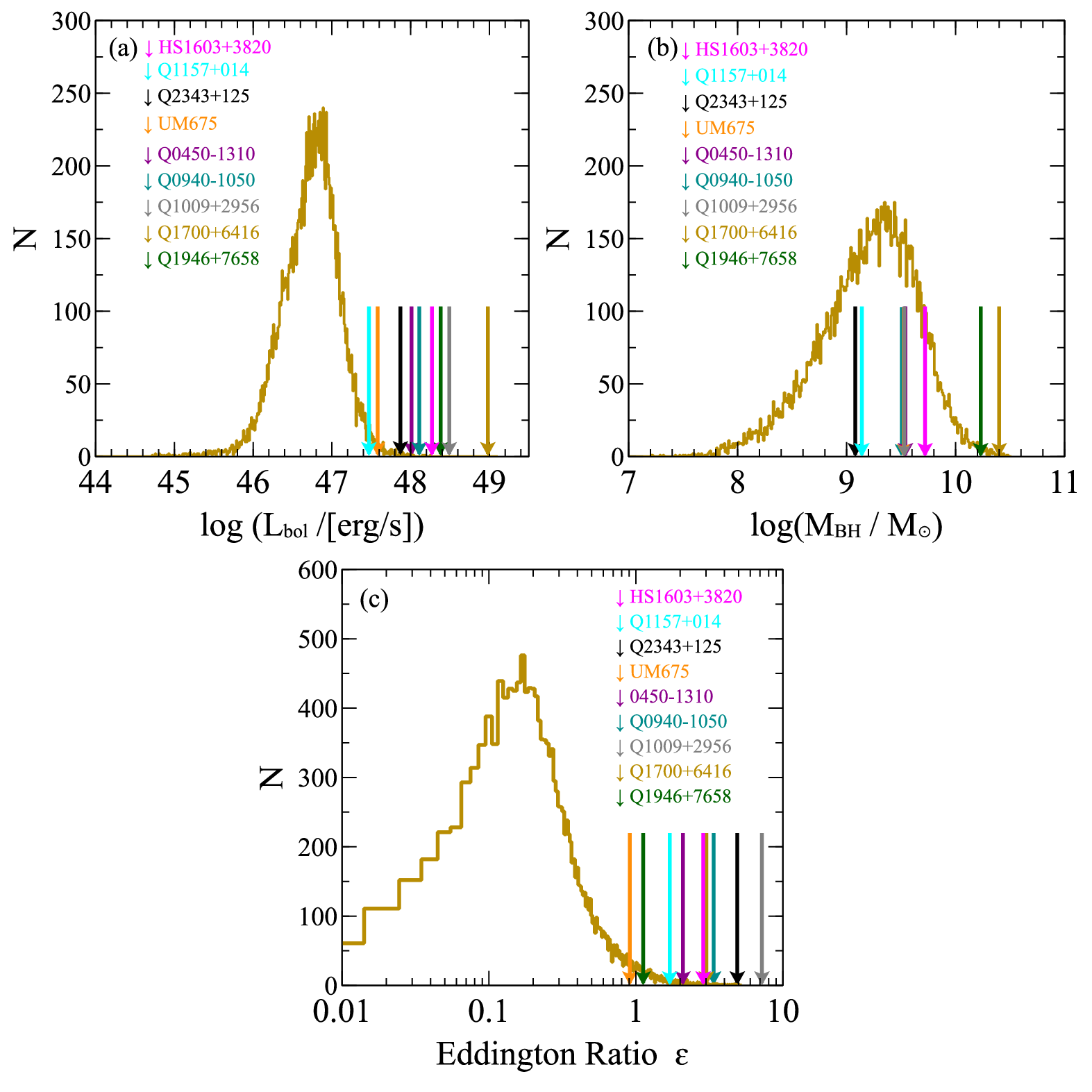}  
 \end{center}
\caption{ (a) Distributions of bolometric luminosity, (b) virial black
  hole mass, and (c) Eddington ratio of our quasars (indicated by
  downward arrows) and $\sim$17,000 SDSS quasars at 2.0 $\leq z <$ 3.1
  \citep{2011ApJS..194...45S} (histograms). Exact values of these
  parameters for our nine quasars are presented in Table 1. }
\label{fig:fig1}
\end{figure*}

\begin{figure*}
 \begin{center}
   \includegraphics[width=14cm]{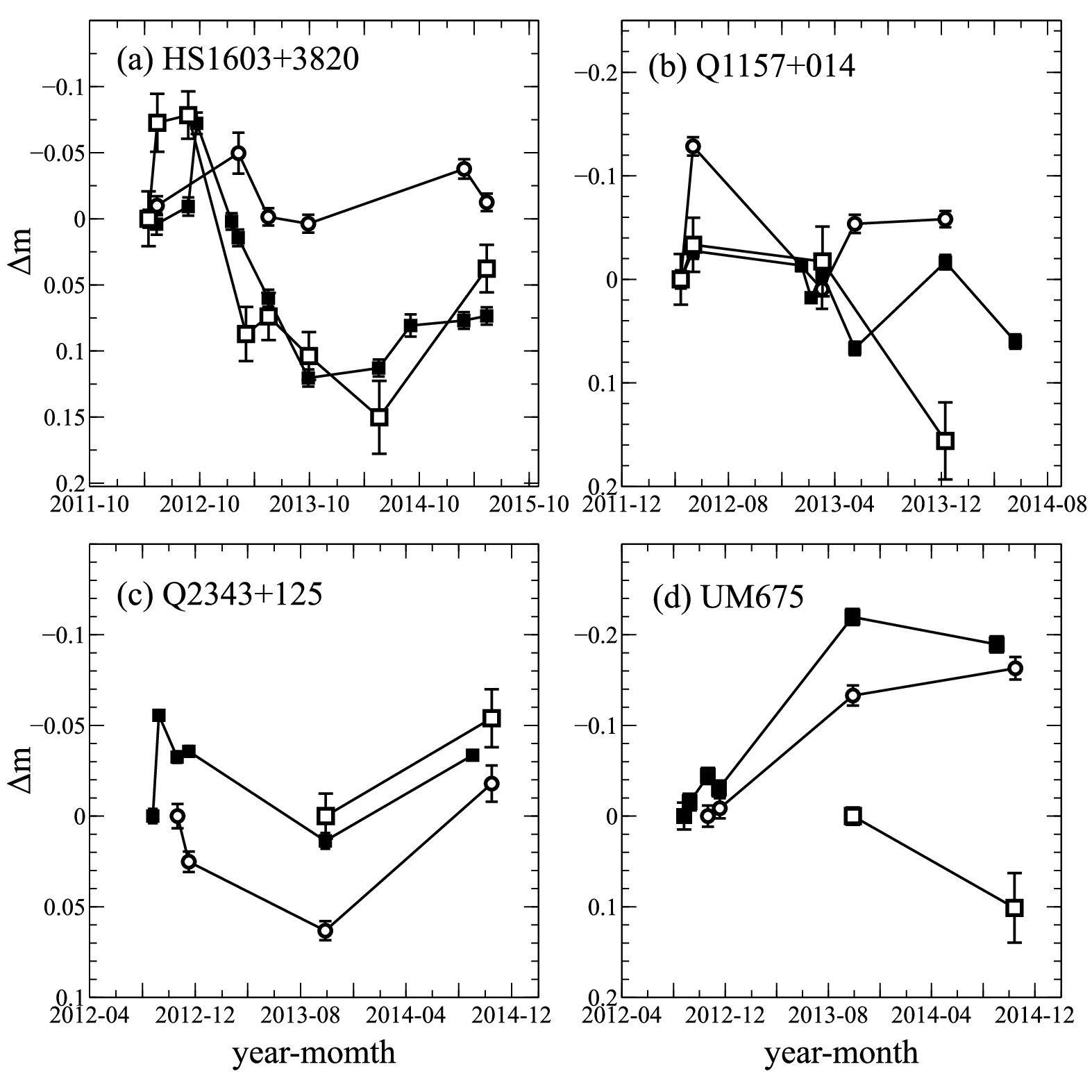} 
 \end{center}
\caption{ Light curves of four mini-BAL quasars ((a) HS1603+3820, (b)
  Q1157+014, (c) Q2343+125, and (d) UM675), monitored in the $u$-band
  ({\it open squares}), $g$ ({\it filled squares}), and $i$-band ({\it
    open circles}). The horizontal axis denotes the observing date
  (year-month) and the vertical axis $\Delta m$ is the magnitude
  difference from the first observation. The $\Delta m$ first
  observing epoch is zero by definition.  }
\label{fig2}
\end{figure*}

\begin{figure*}
 \begin{center}
   \includegraphics[width=16cm]{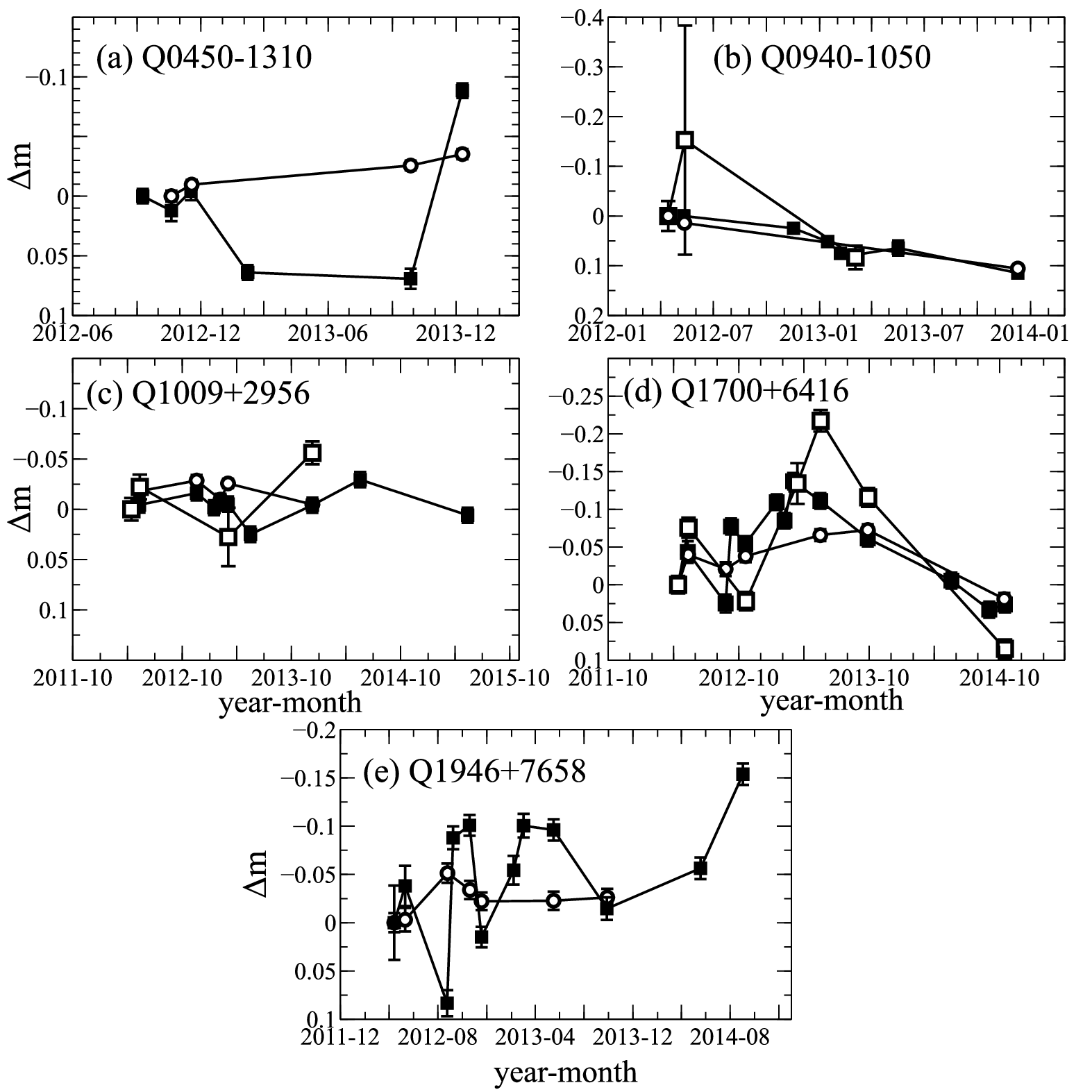} 
 \end{center}
\caption{ Identical to Figure~2, but plotted for the five NAL quasars,
  (a) Q0450-1310, (b) Q0940-1050, (c) Q1009+2956, (d) Q1700+6416, and
  (e) Q1956+7658. }
\label{fig:fig3}
\end{figure*}

\begin{figure*}
\begin{center}
 \includegraphics[width=16cm]{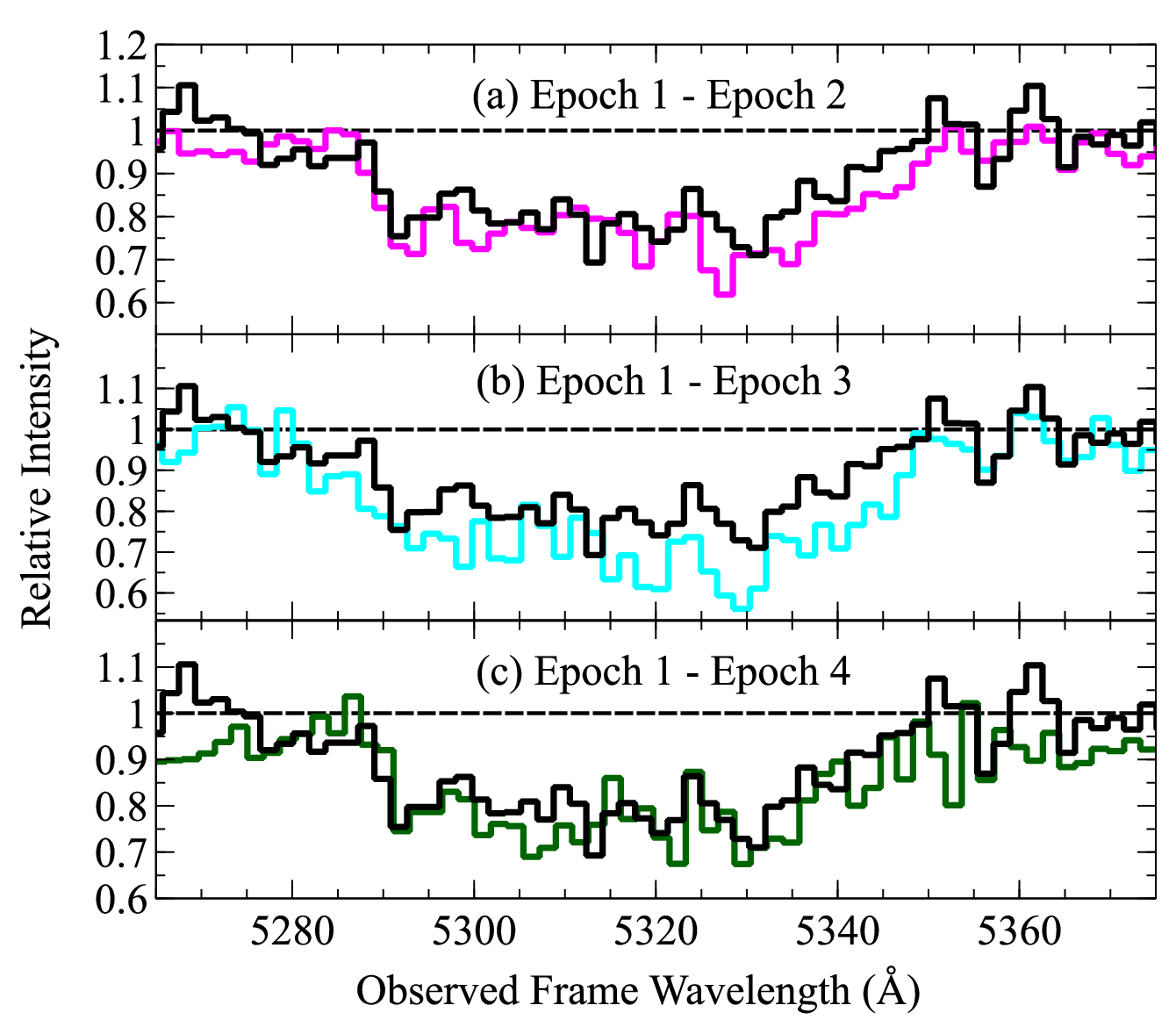} 
\end{center}
\caption{ Normalized spectra of HS1603+3820 around the
  C$\emissiontype{IV}$ mini-BAL in observed frame taken with the
  188-cm Okayama Telescope.  $Black,~magenta,~cyan~ {\rm and}~green$
  histograms denote spectra taken on Sep 19, 2012 (epoch~1), May 30,
  2014 (epoch~2), Feb 23, 2015 (epoch~3), and May 21, 2015 (epoch~4),
  respectively. C$\emissiontype{IV}$ mini-BALs in the (a) epoch~2, (b)
  epoch~3 and (c) epoch~4 are compared to the C$\emissiontype{IV}$
  mini-BAL in epoch~1. Horizontal $dotted$ lines represent the
  normalized continuum levels. }
\label{fig:fig4}
\end{figure*}

\begin{figure*}
 \begin{center}
  \includegraphics[width=12cm, height= 12cm]{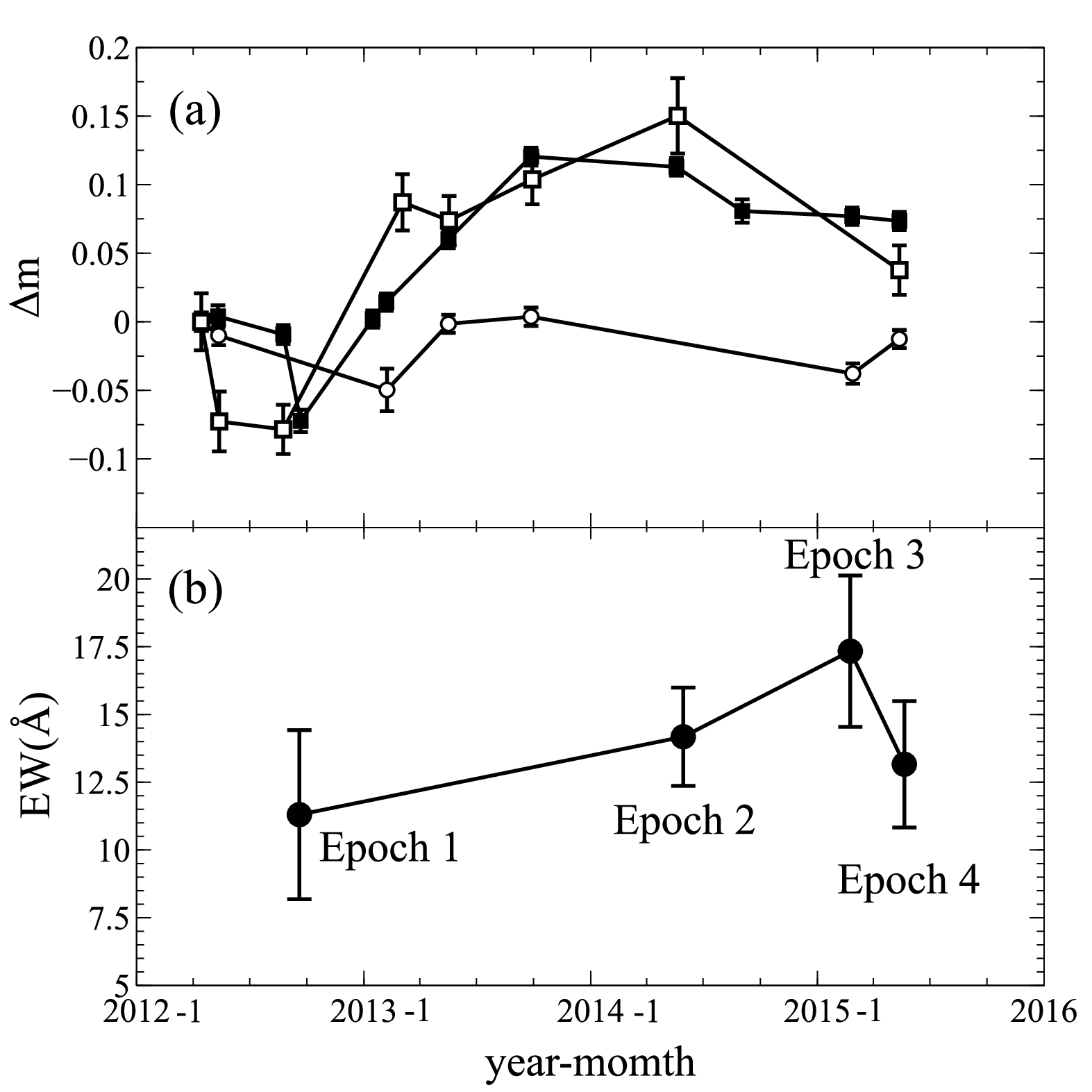} 
  \end{center}
\caption{ (a) Light curves of SDSS $u$-, $g$-, and $i$-band (symbols
  are those of Figures~2 and~3) and (b) the EW variability of
  C$\emissiontype{IV}$ mini-BAL in HS1603+3820.  To clearly compare
  the light curves with the EW variability trend, we invert the
  vertical axis of Figure~2 in this figure. }
\label{fig:fig5}
\end{figure*}

\begin{figure*}
\begin{center}
 \includegraphics[width=12cm]{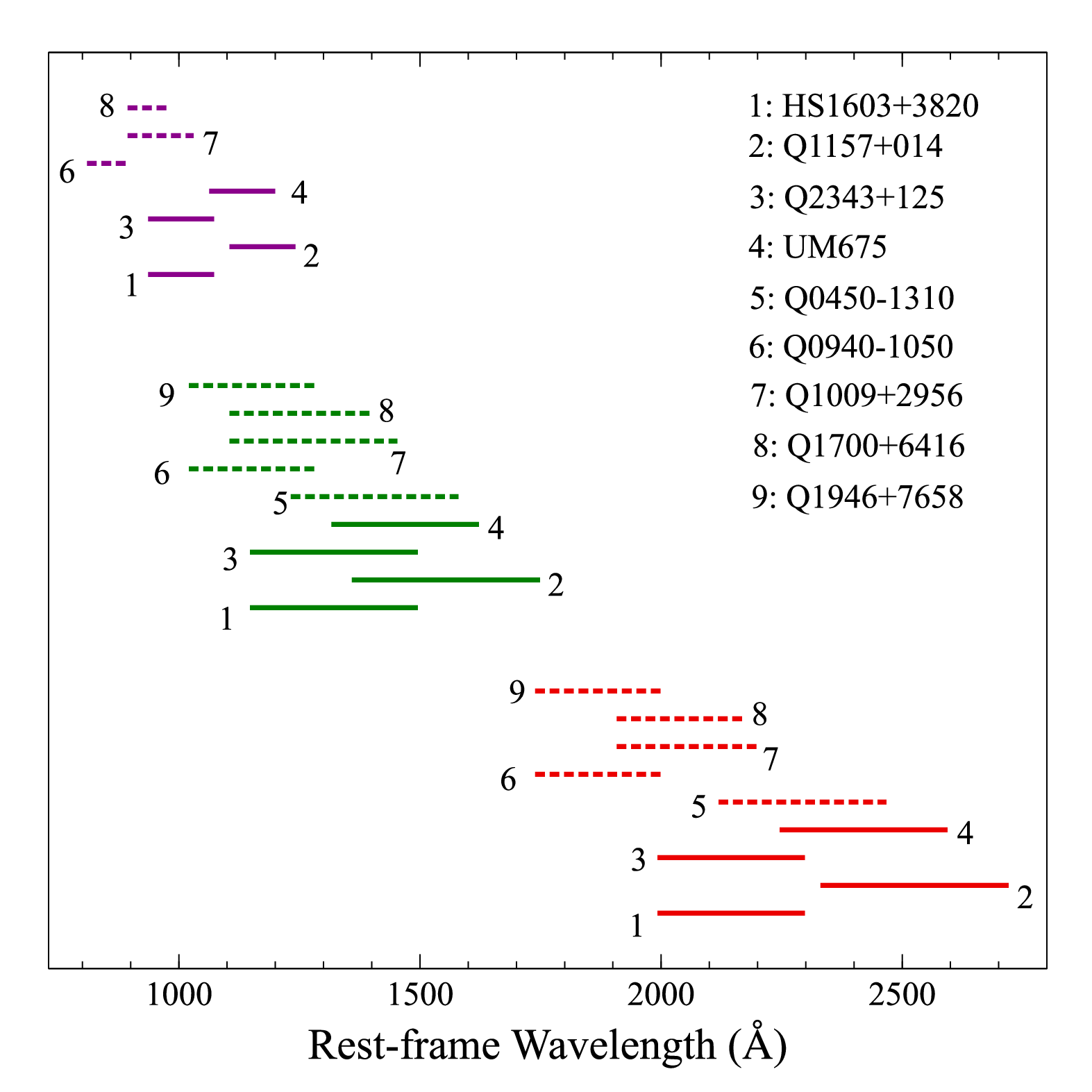} 
\end{center}
\caption{ Regions of rest-frame wavelength covered by SDSS $u$-
  (violet), $g$- (green), and $i$ (red)-band for each quasar. $Solid$
  and $Dotted$ lines represent the wavelength coverage of mini-BAL and
  NAL quasars, respectively. The quasars covering each wavelength
  range are labeled 1~-~9.}
\label{fig:fig6}
\end{figure*}

\begin{figure*}
\begin{center}
   \includegraphics[width=16cm]{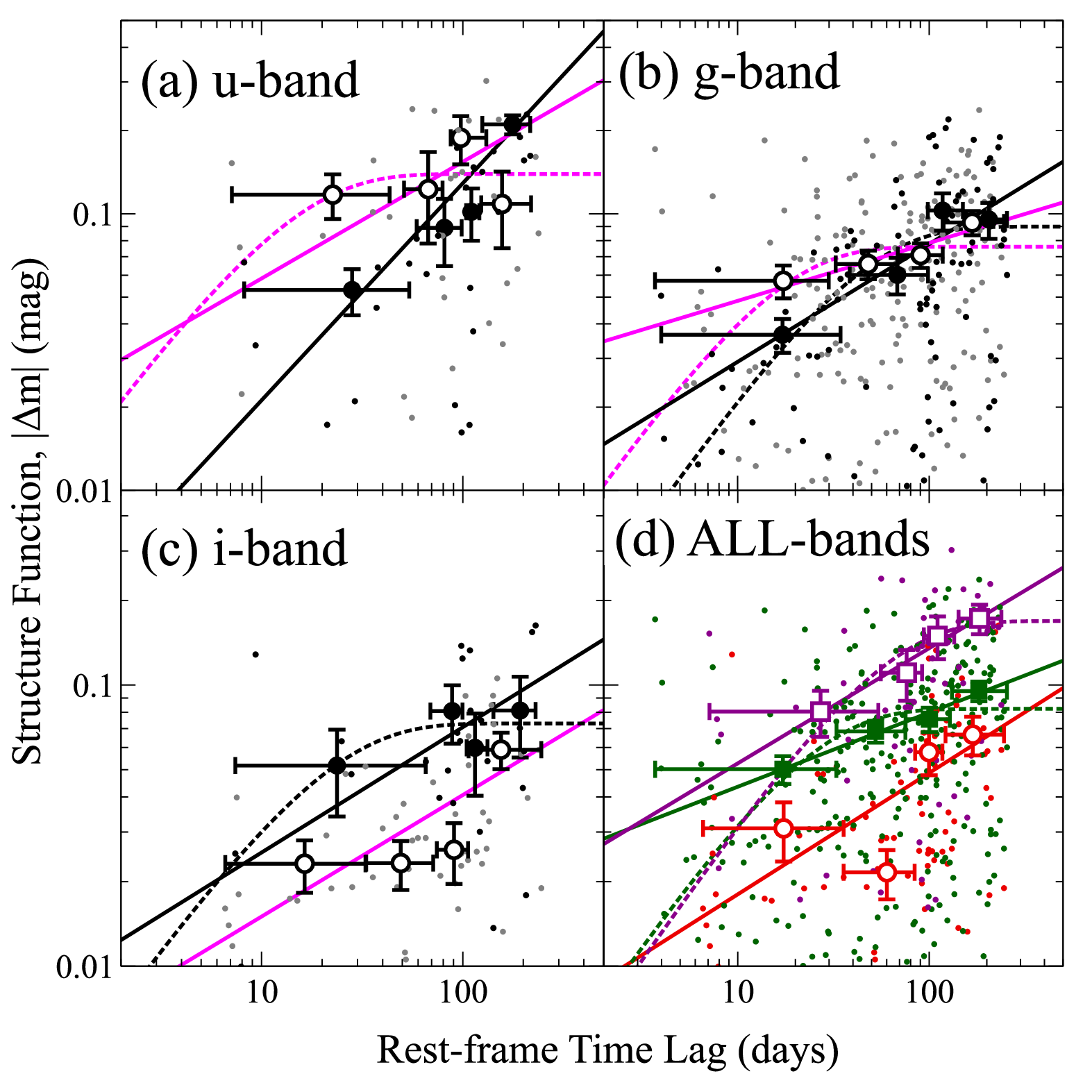}    
\end{center}
\caption{ Structure functions (SFs) of (a) $u$-band, (b) $g$-band ,
  and (c) $i$-band of mini-BAL (filled circles) and NAL (open circles)
  quasars, plotted on a log~-~log scale. The statistical error in the
  SF includes the error propagation.  Horizontal error bars indicate
  the variances from the mean time intervals in each bin.  In panels
  (a), (b), and (c), the quasar variabilities of mini-BAL (black dots)
  and NAL (gray dots) quasars are plotted for all combinations of the
  observing epochs. The SFs of the mini-BAL (black lines) and NAL
  (magenta lines) quasars are fitted by a power low ($solid~line$) and
  an asymptotic function ($dotted~line$), respectively. (d) The SFs of
  all subsamples including mini-BAL and NAL quasars in the $u$-band
  (violet), $g$-band (green) and $i$-band (red) are also fitted to
  power-law and asymptotic functions. The quasar variabilities of all
  our quasars in $u$-band (violet dots), $g$-band (green dots), and
  $i$-band (red dots) are also plotted for all combinations of the
  observing epochs. Unsatisfactory fitting results are omitted. }
\label{fig:fig7}
\end{figure*}

\begin{figure*}
\begin{center}
   \includegraphics[width=12cm]{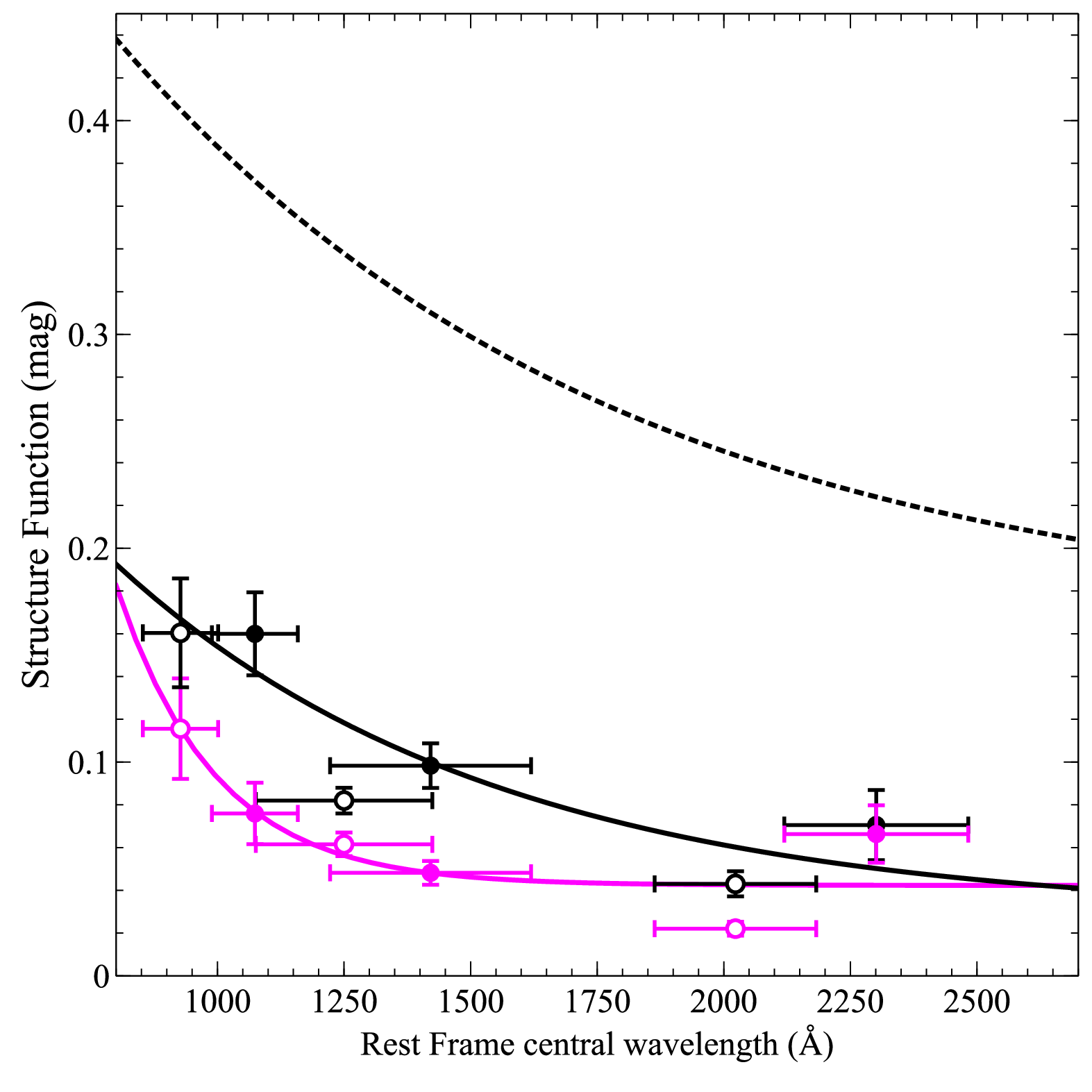} 
\end{center}
\caption{ Structure function versus rest-frame wavelength. The samples
  were first separated into two subsamples with longer and shorter
  time-lags. The separation criterion was $\Delta \tau$ = 90~days in
  the rest-frame. Filled and open circles (magenta: $\Delta
  \tau~<~$90, black: $\Delta \tau~>~$90) indicate the SFs of mini-BAL
  and NAL quasars, respectively. For each mini-BAL / NAL quasar, the
  rest-frame central wavelength denotes the average central
  (rest-frame) wavelengths among all bands. Horizontal error bars
  indicate the bandwidth of each filter. {\it Solid} black ($\Delta
  \tau~>~$90) and magenta ($\Delta \tau~<~$90) curves are the fitting
  results. Black $dotted$ curve is fitted to the $\sim$25000 normal
  quasars from VB04 data by Eq. (7) ($A~=~0.616~\pm~0.056,
  ~\lambda_0~=~988~\pm~60, ~B~=~0.164~\pm~ 0.003$).  }
\label{fig:fig8}
\end{figure*}

\begin{figure*}
 \begin{center}
  \includegraphics[width=16cm]{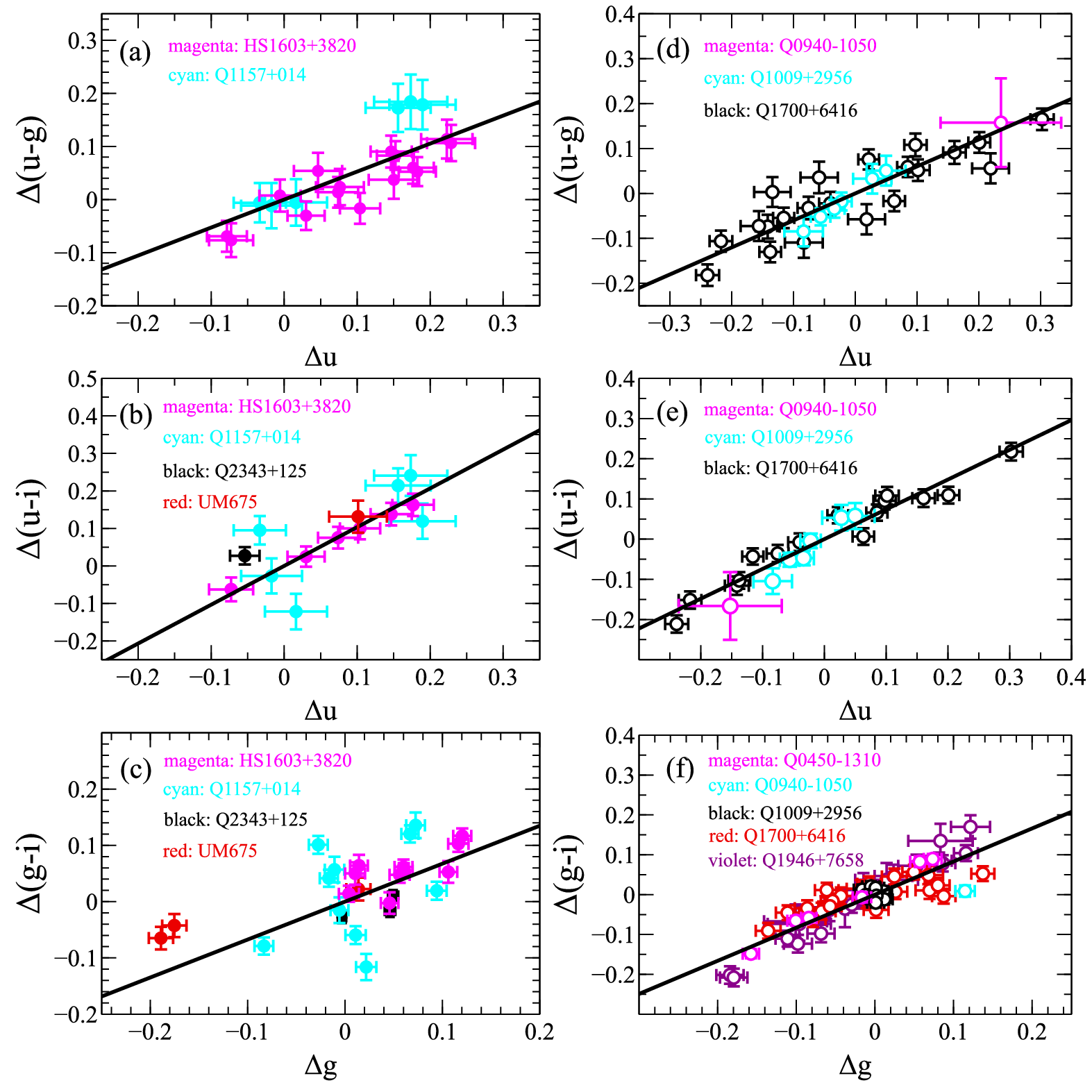} 
 \end{center}
\caption{ Color variability of $\Delta (u-g)$ (top), $\Delta (u-i)$
  (middle) and $\Delta (g-i)$ (bottom) versus magnitude variability in
  mini-BAL (left column: (a), (b), and (c)) and NAL (right column:
  (d), (e), and (f)) quasars. Magnitude variabilities were determined
  in the bluer bands. Solid lines are the best fits to the
  distributions. }
\label{fig:fig9}
\end{figure*}

\begin{landscape}
\begin{table}
\begin{center}
\tbl{Sample Quasars}{%
\label{table:data_type}
\begin{tabular}
{
@{\hspace{0.00cm}}
c@{\hspace{0.15cm}}
c@{\hspace{0.15cm}}
c
c
c@{\hspace{0.15cm}}
c@{\hspace{0.2cm}}
c@{\hspace{0.1cm}}
c@{\hspace{0.4cm}}
c@{\hspace{0.2cm}}
c@{\hspace{0.2cm}}
c@{\hspace{0.2cm}}
c@{\hspace{0.2cm}}
c@{\hspace{0.1cm}}
c
c
c
}
\hline
Quasar & 
RA$^{*}$ & 
Dec$^{\dagger}$ & 
$m_{\rm V}$ ($M_{\rm V}$)$^{\ddagger}$&
$z_{\rm em}^{\S}$ & 
$z_{\rm abs}^{\|}$ & 
$v_{\rm ej}^{\#}$ & 
Variability$^{**}$ & 
$\langle {\rm EW}_{\rm abs, C\emissiontype{IV}} \rangle ^{\dagger \dagger}$ & 
$\langle \Delta {\rm EW} \rangle ^{\ddagger \ddagger}$ &
$(\Delta {\rm EW})_{\rm max}^{\S \S}$&
$R$$^{\| \|}$ & 
$\log L_{\rm bol}^{\# \#}$ & 
$\log$$M_{\rm BH}$/$M_{\solar}$$^{***}$ & 
$\varepsilon^{\dagger \dagger \dagger}$ &
Ref. $^{\ddagger \ddagger \ddagger}$\\
&
&
&
(mag)&
&
&
(km s$^{-1}$)&
&
(${\rm \AA}$)&
&
&
&
&
&\\
\hline 
\multicolumn{16}{c}{mini-BAL Quasar}\\
\hline	
HS1603+3820 & 16:04:55.4  & +38:12:01 & 15.99 ($-$30.60) &2.542 & $\sim$2.43 &  $\sim$9500  & Y & 13.10 &2.03$\pm$0.38 & 7.83$\pm$2.16 & $<$0.2 & 48.27 & 9.72 & 2.87 & 1,5,8 \\
Q1157+014  & 11:59:44.8  & +01:12:07  & 17.52 ($-$28.49) &2.00  & $\sim$1.97 &  $\sim$3000  & Y$^{\S \S \S}$ & 37.96 & 1.09$\pm$1.21 &1.41$\pm$1.61 & 471 & 47.47 
& 9.14 & 1.70 & 2,5,2  \\
Q2343+125   & 23:46:28.2  &  +12:49:00  &17.0  ($-$29.62) & 2.515 & $\sim$2.24 &  $\sim$24400 & N & 2.48  
&0.84$\pm$0.48 &1.25$\pm$0.82 & 1.27  & 47.87 & 9.08 & 4.90 & 11,5,9 \\  
UM675 & 01:52:27.3  &$-$20:01:06  & 17.4  ($-$28.81) &2.15  & $\sim$2.13 & $\sim$1900  & Y &  4.51 & ---$^{\l \l \l}$ & 1.54$\pm$0.32 & 438  & 47.58 & 9.52 & 0.91 & 3,5,10 \\
\hline
\multicolumn{16}{c}{NAL Quasar}\\
\hline
Q0450-1310  & 04:53:13.6  &$-$13:05:55  & 16.5 ($-$29.89) & 2.300 & 2.2307 & 37037 & N & --- & --- & --- & $<$1.69 & 48.01 & 9.59  & 1.90 & 4,12,12 \\
Q0940-1050  & 09:42:53.4  &$-$11:04:25  & 16.90 ($-$30.26) & 3.080 & 2.8347 &  18578 & N & 1.64 & 0.03$\pm$0.04 & 0.04$\pm$0.06 & $<$2.58 & 48.11 & 9.48  & 3.59 & 4,12,12 \\
Q1009+2956 & 10:11:56.6  & +29:41:41 & 16.05 ($-$30.71) & 2.644 & 2.2533 &  33879 & N & 1.73  & ---$^{\l \l \l}$ & 0.01$\pm$0.07 & $<$1.58 & 48.49 & 9.53  & 7.21 & 4,6,8 \\  
Q1700+6416  & 17:01:00.6  & +64:12:09  & 16.17 ($-$30.66) & 2.722 & 2.7125 &  767 & N & 0.30 & 0.02$\pm$0.01 & 0.03$\pm$0.02 & $<$1.24 & 48.98 & 10.4  & 3.02 & 4,6,2 \\
Q1946+7658  & 19:44:55.0  &  +77:05:52  &  16.20 ($-$30.94) & 3.051 & 2.8928 &  927 & N & 0.29 & ---$^{\# \# \#}$ & ---$^{\# \# \#}$ & $<$1.35 & 48.38 & 10.23 & 1.12 & 4,7,7 \\
\hline
\multicolumn{16}{@{}l@{}}{\hbox to 0pt{\parbox{180mm}{\footnotesize 
Notes ---
\footnotemark[*] Right Ascension.
\footnotemark[$\dagger$] Declination.
\footnotemark[$\ddagger$] V-band magnitude (Vega) from
\citet{2010A&A...518A..10V}.  Values in parentheses are absolute
magnitudes.
\footnotemark[$\S$] C$\emissiontype{IV}$ emission redshift.
\footnotemark[$\|$] Apparent redshift of C$\emissiontype{IV}$ outflow.
\footnotemark[$\#$] Ejection velocity determined from the quasar
emission redshift (in km s$^{-1}$).
\footnotemark[**] Absorption line variability (Yes or No). See
\citet{2014ApJ...792...77M}.
\footnotemark[$\dagger \dagger$] Averaged equivalent width of
C$\emissiontype{IV}$ absorption line given by the outflows (in \AA),
from \citet{2014ApJ...792...77M}.
\footnotemark[$\ddagger \ddagger$] Averaged amplitude of
  C$\emissiontype{IV}$ absorption variabilities, from
  \citet{2014ApJ...792...77M}.
\footnotemark[$\S \S$] Maximum amplitude of C$\emissiontype{IV}$
  absorption variabilities, from \citet{2014ApJ...792...77M}.
\footnotemark[$\| \|$] Radio loudness.
\footnotemark[$\# \#$] Bolometric luminosity. 
\footnotemark[***] Central black hole mass (in units of solar units).
\footnotemark[$\dagger$ $\dagger$ $\dagger$] Eddington ratio, $L_{\rm
  bol}$/$L_{\rm Edd}$.
\footnotemark[$\ddagger$ $\ddagger$ $\ddagger$] References for $R$,
$\log L_{bol}$, and $\log M_{BH}$ in numerical order --- (1)
\citet{2007ApJ...665.1004J}, (2) \citet{2011ApJS..194...45S}, (3)
\citet{1994ApJS...90..179G}, (4) \citet{2007ApJS..171....1M}, (5)
\citet{2014ApJ...792...77M}, (6) \citet{2010ApJ...722..997W}, (7)
\citet{1995ApJ...438..643K}, \citet{2014NewA...28...70R}, (9)
\citet{2012ApJ...752...39T}, (10) \citet{2009ApJ...696.1998D}, (11)
FIRST survey, and (12) This paper.
\footnotemark[$\S \S \S$] Variability is seen only in
  Si$\emissiontype{IV}$ mini-BAL with a significance level of
  $\sim$~2.4$\sigma$ \citep{2014ApJ...792...77M}.
\footnotemark[$\l \l \l$] {Cannot be calculated because our sample
  was limited to two epochs.}
\footnotemark[$^{\# \# \#}$] We cannot calculate these because
C$\emissiontype{IV}$ NAL was observed only once
\citep{2014ApJ...792...77M}.}  } \hss }
\end{tabular}}
\end{center}
\end{table}
\end{landscape}

\begin{longtable}{ccccc}
\caption{Log of observations}
\hline 
QSO & Obs-Date & Band & $\Delta t_{\rm rest}^{*}$ & $t_{\rm EXP}^{\dagger}$ \\
     & (day)  &      & (s)            \\
\hline
\endhead
\hline 
\endfoot
\hline
HS1603+3820 (mini-BAL QSO)& 2012 Apr 14& $u$&0& 180$\times$5\\
&2012 Apr 14&$g$&0&60$\times$5\\
&2012 Apr 14&$i$&0& 60$\times$5\\
&2012 May 12&$i$&7.9& 60$\times$5\\
&2012 May 12&$g$&8.2& 60$\times$5\\
&2012 May 13&$u$&8.2& 300$\times$5\\
&2012 Aug 24&$u$&37.3&300$\times$5\\
&2012 Aug 24&$g$&37.3&60$\times$3\\
&2012 Sep 21& $g$&45.2& 180$\times$3\\
&2013 Jan 15&$g$& 77.9 & 180$\times$5 \\
&2013 Feb 6&$g$& 84.1 &60$\times$5\\
&2013 Feb 7&$i$& 84.4 &300$\times$3\\
&2013 Mar 4& $u$& 91.5 & 300$\times$3\\
&2013 May 17&$g$& 112.4 &60$\times$5\\ 
&2013 May 17&$i$& 112.4 &60$\times$5\\
&2013 May 18&$u$& 112.6 &420$\times$1, 480$\times$3, 600$\times$1\\
&2013 Sep 27&$g$& 149.9 &120$\times$5\\
&2013 Sep 27&$i$& 149.9 &120$\times$5\\
&2013 Sep 29&$u$& 150.5 & 300$\times$5 \\
&2014 May 19&$g$& 215.0 &60$\times$5\\
&2014 May 21&$u$& 216.5 & 300$\times$4\\ 
&2014 Sep 2&$g$& 245.9 &120$\times$2, 180$\times$2, 240$\times$1\\ 
\hline
Q1157+014 (mini-BAL QSO)& 2012 Apr 14& $u$& 0 & 300$\times$5\\
&2012 Apr 14&$g$& 0 & 120$\times$5\\
&2012 Apr 14&$i$& 0 & 60$\times$1, 120$\times$4\\
&2012 May 12&$u$& 9.3 &300$\times$5\\
&2012 May 12&$g$& 9.3 & 120$\times$5\\
&2012 May 12&$i$& 9.3 &120$\times$5\\
&2013 Jan 15&$g$& 92 & 180$\times$1, 300$\times$5\\
&2013 Feb 6&$g$& 99.3 &180$\times$5\\
&2013 Mar 3&$g$& 107.7 &180$\times$5\\
&2013 Mar 3&$i$& 107.7 &180$\times$3\\
&2013 Mar 4&$u$& 108.0 & 600$\times$3\\
&2013 May 17&$g$& 132.7 &120$\times$5\\
&2013 May 17&$i$& 132.7 &120$\times$5\\
&2013 Dec 10&$u$& 201.7 & 600$\times$4\\
&2013 Dec 10&$g$& 201.7 &360$\times$3\\ 
&2013 Dec 10&$i$& 201.7 &180$\times$5\\
&2014 May 19&$g$& 255.0 &120$\times$5\\
\hline
Q2343+125 (mini-BAL QSO) &2012 Aug 25& $g$& 0 &120$\times$1, 180$\times$1, 240$\times$1\\
&2012 Sep 8&$g$&4.0& 120$\times$5\\
&2012 Oct 21&$g$&16.2& 120$\times$5\\
&2012 Oct 21&$i$& 0 & 120$\times$5\\
&2012 Nob 16&$g$& 23.6 & 120$\times$4\\
&2012 Nob 16&$i$& 7.4 & 120$\times$5 \\
&2013 Sep 27&$g$& 113.2 &120$\times$4, 240$\times$1\\
&2013 Sep 27&$i$& 97.0 &120$\times$5\\
&2013 Sep 28&$u$& 0 &300$\times$5\\
&2014 Sep 2&$g$& 209.9 &120$\times$1, 180$\times$4\\
&2014 Oct 16& $u$&109.0&300$\times$1, 360$\times$4\\
&2014 Oct 16& $g$& 222.5 &120$\times$5\\ 
&2014 Oct 16& $i$& 206.2 &120$\times$4\\ 
\hline
UM675 (mini-BAL QSO)&2012 Aug 26& $g$& 0 &300$\times$2\\
&2012 Sep 8& $g$& 4.1 &120$\times$5\\
&2012 Oct 21& $g$& 17.8 & 120$\times$5 \\
&2012 Oct 21& $i$& 0 & 120$\times$5\\
&2012 Nob 17& $g$& 26.0 & 120$\times$4 \\
&2012 Nob 18& $i$& 8.9 &120$\times$5\\
&2013 Sep 27& $g$& 126.0 &180$\times$1, 240$\times$3\\
&2013 Sep 28& $u$& 0 &420$\times$5\\
&2013 Sep 28& $i$& 108.6 &120$\times$5\\
&2014 Sep 2& $g$& 234.0 &180$\times$3\\
&2014 Oct 16& $u$& 121.6 & 300$\times$2\\
&2014 Oct 16& $i$& 230.2 &120$\times$4\\ 
\hline
Q0450-1310 (NAL QSO)& 2012 Sep 9& $g$& 0 & 60$\times$2, 120$\times$3\\
&2012 Oct 20& $g$& 12.4 & 60$\times$3, 120$\times$2\\
&2012 Oct 20& $i$&0& 120$\times$5\\
&2012 Nob 17& $g$& 20.9 &180$\times$3 \\
&2012 Nob 18& $i$& 8.8 & 60$\times$5\\
&2013 Feb 6& $g$& 45.4 & 180$\times$2, 240$\times$1\\
&2013 Sep 27& $g$& 116.0 &120$\times$5\\
&2013 Sep 27& $i$&103.6 &60$\times$5\\
&2013 Dec 10& $g$&138.5 &240$\times$5\\
&2013 Dec 10& $i$&126.1 &60$\times$5\\ 
\hline
Q0940-1050 (NAL QSO)& 2012 Apr 14& $u$& 0 & 300$\times$5\\
&2012 Apr 14& $i$& 0 & 60$\times$5\\
&2012 May 11& $g$& 0 &60$\times$5\\
&2012 May 12& $i$& 6.9 & 60$\times$5\\
&2012 May 13& $u$& 7.1 & 300$\times$3\\
&2012 Nob 17& $g$&46.6&300$\times$4, 240$\times$1\\
&2013 Jan 15& $g$& 61.0 &180$\times$5\\
&2013 Feb 6& $g$& 66.4 &180$\times$5\\
&2013 Mar 3& $g$& 72.5 &120$\times$5\\
&2013 Mar 4& $u$& 79.4 &  600$\times$2\\
&2013 May 17& $g$& 90.9 &60$\times$5\\
&2013 Dec 10& $g$&142.0&180$\times$5\\
&2013 Dec 10& $i$& 148.3 &120$\times$5\\
\hline
Q1009+2956 (NAL QSO)& 2012 Apr 14& $u$& 0 & 300$\times$5\\
&2012 Apr 14& $g$& 0 & 60$\times$5\\
&2012 Apr 14& $i$& 0 & 60$\times$4\\
&2012 May 11& $u$& 8.0 & 300$\times$5\\
&2012 May 11& $g$&7.4& 60$\times$5\\
&2012 May 12& $i$&7.7&60$\times$5\\
&2012 Nob 18& $g$& 59.8 &  180$\times$1, 300$\times$4 \\
&2012 Nob 18& $i$& 59.8 &120$\times$5\\
&2013 Jan 15& $g$& 75.7 & 180$\times$6\\
&2013 Feb 6& $g$& 81.8 &60$\times$4, 180$\times$1\\
&2013 Feb 7& $i$& 82.0 &300$\times$5\\
&2013 Mar 3& $g$& 88.6 &60$\times$4, 120$\times$1\\
&2013 Mar 3& $i$& 88.6 &60$\times$4, 120$\times$1\\
&2013 Mar 4& $u$& 88.9& 300$\times$3 \\
&2013 May 17& $g$& 109.2 &60$\times$5\\
&2013 Dec 10& $u$& 166.0 & 300$\times$5 \\
&2013 Dec 10& $g$& 166.0 &120$\times$5\\
&2013 Dec 10& $i$& 166.0 &60$\times$5\\
&2014 May 19& $g$& 209.9 &60$\times$5\\
\hline
Q1700+6416 (NAL QSO)& 2012 Apr 14& $u$& 0 &180$\times$5\\
&2012 Apr 14& $g$& 0 & 60$\times$5\\
&2012 Apr 14& $i$& 0 & 60$\times$5\\
&2012 May 11& $g$&7.2&60$\times$5\\
&2012 May 12& $i$& 7.5 & 60$\times$5\\
&2012 May 13& $u$& 7.8 & 300$\times$5\\
&2012 Aug 25& $g$& 35.7 &120$\times$1, 240$\times$1, 300$\times$1\\
&2012 Aug 25& $i$& 35.7  & 300$\times$3 \\
&2012 Sep 9& $g$& 39.8 & 180$\times$2, 300$\times$1\\
&2012 Oct 19& $g$& 50.5 &60$\times$5\\
&2012 Oct 20& $i$& 50.8 & 60$\times$5 \\
&2012 Oct 21& $u$& 51.0 &300$\times$5\\
&2013 Jan 15& $g$& 74.2 & 180$\times$5\\
&2013 Feb 6& $g$& 80.1 & 60$\times$5\\
&2013 Mar 3& $g$& 86.5 &180$\times$5\\
&2013 Mar 4& $u$& 87.0 & 300$\times$5 \\
&2013 May 17& $g$& 106.9 &60$\times$5\\
&2013 May 17& $i$& 106.9 &60$\times$5\\
&2013 May 18& $u$& 107.2 & 300$\times$1, 480$\times$1, 600$\times$2\\
&2013 Sep 27& $g$& 142.7 &120$\times$5\\
&2013 Sep 27& $i$&142.7&60$\times$4\\
&2013 Sep 28& $u$& 142.9 & 300$\times$5 \\
&2014 May 19& $g$& 205.5 &60$\times$5\\
&2014 Sep 2& $g$& 234.0 &120$\times$5\\
&2014 Oct 16& $u$& 238.0 & 300$\times$5\\ 
&2014 Oct 16& $g$& 245.8 &60$\times$4, 120$\times$1\\
&2014 Oct 16& $i$& 245.0 & 60$\times$1, 120$\times$3\\ 
\hline
Q1946+7658 (NAL QSO)& 2012 Apr 14& $g$& 0 & 60$\times$5\\
& 2012 Apr 14& $i$& 0 & 60$\times$5\\
&2012 May 11& $g$&6.7&60$\times$5\\
&2012 May 11& $i$& 6.7 & 60$\times$5\\
&2012 Aug 24& $g$& 32.6 &300$\times$3\\
&2012 Aug 25& $i$& 32.8 & 120$\times$1, 300$\times$2\\
&2012 Sep 8& $g$& 36.3 & 60$\times$2, 120$\times$3\\
&2012 Oct 19& $g$& 46.4  & 60$\times$6\\
&2012 Oct 20& $i$& 46.6 &60$\times$5\\
&2013 Nob 18& $g$&53.8 &120$\times$5\\
&2012 Nob 18& $i$& 53.8 &120$\times$5\\
&2013 Feb 6& $g$& 73.6&60$\times$1, 300$\times$2\\
&2013 Mar 3& $g$& 79.7 &60$\times$5\\
&2013 May 17& $g$& 98.2 &60$\times$5 \\
&2013 May 17& $i$& 98.2 & 60$\times$5\\
&2013 Sep 27& $g$& 124.4 &120$\times$1, 180$\times$2, 240$\times$1\\
&2013 Sep 28& $i$& 131.3 &60$\times$5\\ 
&2014 May 19& $g$& 188.8 &60$\times$5\\
&2014 Sep 2& $g$& 215.0 &60$\times$1, 120$\times$4\\
\hline
\multicolumn{5}{@{}l@{}}{\hbox to 0pt{\parbox{180mm}{\footnotesize
Notes ---
\footnotemark[*] Time delay from the first observation in the quasar
rest-frame.  Zero denotes the first epoch.\\
\footnotemark[$\dagger$] Total exposure time for usable image, which
is altered according to the weather.\\
}\hss}}
\endlastfoot
\end{longtable}

\begin{table*}
\begin{center}
 \tbl{Spectroscopic observation log of HS1603+3820}{%
\begin{tabular}{ccc}
\hline
Observing Epoch&
Obs-Date &
$t_{\rm EXP}^{*}$ \\
 & (day) & (s) \\
\hline
 1&2012 Sep 19&1,200$\times$2\\
 2&2014 May 30&1,200$\times$8\\
 3&2015 Feb 23&1,200$\times$3\\
 4&2015 May 21&1,200$\times$3\\
\hline
\multicolumn{3}{@{}l@{}}{\hbox to 0pt{\parbox{85mm}{\footnotesize
\footnotemark[*] Total exposure time for usable image.\\
}\hss}}
\end{tabular}}
\end{center}
\end{table*}

\begin{table*}
\begin{center}
 \tbl{Detailed variability properties of the light curves of min i-BAL and NAL quasars}{%
 \label{table:tab3}
 \begin{tabular}{cccccccc}
\hline
Quasar&
Type&
$N^{*}$&
$\sigma_m$$^{\dagger}$ &
 $\langle |\Delta m| \rangle^{\ddagger}$ &
 $|\Delta m_{\rm max}|^{\S}$ &
 $\langle | \Delta m/\Delta t_{\rm rest}| \rangle$$^{\|}$&
$|\Delta m/\Delta t_{\rm rest}|_{\rm max}$$^{\#}$\\
&
&
&
(mag)&
(mag)&
(mag)&
(mag/yr)&
(mag/yr)\\
\hline
\multicolumn{8}{c}{SDSS $u$-band}\\
\hline
HS1603+3820&mini-BAL QSO&7&0.068&0.104$\pm$0.015&0.229$\pm$0.035&0.387$\pm$ 0.040&1.116$\pm$0.204\\
Q1157+014&mini-BAL QSO&4&0.084&0.086$\pm$0.033$\dagger \dagger$&0.189$\pm$0.045&0.285$\pm$0.070&0.676$\pm$0.196\\
Q2343+125&mini-BAL QSO&2&--- $^{**}$&--- $^{**}$&0.054$\pm$0.020$^{\dagger \dagger}$&--- $^{**}$&0.181$\pm$0.068$^{\dagger \dagger}$\\
UM675&mini-BAL QSO&2&---$^{**}$&---$^{**}$&0.101$\pm$0.040$^{\dagger \dagger}$&---$^{**}$&0.304$\pm$0.119$^{\dagger \dagger}$\\
Q0940-1050&NAL QSO&3&0.080&0.138$\pm$0.042&0.236$\pm$0.098$^{\dagger \dagger}$&0.634$\pm$0.402&1.191$\pm$0.496$^{\dagger \dagger}$\\
Q1009+2956&NAL QSO&4&0.023&0.041$\pm$0.008&0.056$\pm$0.016&0.116$\pm$0.028&0.123$\pm$0.035\\
Q1700+6416&NAL QSO&7&0.076&0.128$\pm$0.017&0.302$\pm$0.019&0.326$\pm$0.063&3.546$\pm$0.831\\
\hline
\multicolumn{8}{c}{SDSS $g$-band}\\
\hline
HS1603+3820&mini-BAL QSO&10&0.049&0.069$\pm$0.007&0.193$\pm$0.009&0.229 $\pm$0.021&2.909$\pm$0.451\\
Q1157+014&mini-BAL QSO&8&0.030&0.040$\pm$0.005&0.094$\pm$0.010&0.109$\pm$0.020&1.549$\pm$0.358\\
Q2343+125&mini-BAL QSO&7&0.020&0.023$\pm$0.004&0.067$\pm$0.001&0.042$\pm$0.014&4.634$\pm$0.058\\
UM675&mini-BAL QSO&6&0.083&0.110$\pm$0.021&0.220$\pm$0.017&0.334$\pm$0.041&0.691$\pm$0.045\\
Q0450-1310&NAL QSO&6&0.047&0.070$\pm$0.012&0.158$\pm$0.010&0.301$\pm$0.056&2.568$\pm$0.169\\
Q0940-1050&NAL QSO&7&0.028&0.046$\pm$0.006&0.115$\pm$0.012&0.304$\pm$0.023&0.929$\pm$0.196\\
Q1009+2956&NAL QSO&9&0.014&0.015$\pm$0.002&0.054$\pm$0.011&0.052$\pm$0.009&$0.303\pm$0.080\\
Q1700+6416&NAL QSO&13&0.044&0.069$\pm$0.005&0.170$\pm$0.016&0.193$\pm$0.019&9.248$\pm$1.545\\
Q1946+7658&NAL QSO&12&0.052&0.076$\pm$0.007&0.237$\pm$0.017&0.249$\pm$0.033&16.900$\pm$1.778\\
\hline
\multicolumn{8}{c}{SDSS $i$-band}\\
\hline
HS1603+3820&mini-BAL QSO&5&0.024&0.012$\pm$0.005$\dagger \dagger$&0.053$\pm$0.016&0.033$\pm$0.020$\dagger \dagger$&0.630$\pm$0.208\\
Q1157+014&mini-BAL QSO&5&0.065&0.065$\pm$0.013&0.138$\pm$0.021&0.145$\pm$0.050$\dagger \dagger$&5.027$\pm$0.4887\\
Q2343+125&mini-BAL QSO&2&0.024&0.044$\pm$0.009&0.081$\pm$0.011&0.117$\pm$0.040$\dagger \dagger$&0.238$\pm$0.032\\
UM675&mini-BAL QSO&4&0.066&0.102$\pm$0.027&0.163$\pm$0.017&0.273$\pm$0.043&0.456$\pm$0.058\\
Q0450-1310&NAL QSO&4&0.010&0.020$\pm$0.004&0.035$\pm$0.005&0.086$\pm$0.009&0.090$\pm$0.019\\
Q0940-1050&NAL QSO&3&0.052&0.083$\pm$0.025&0.105$\pm$0.007&0.254$\pm$0.011&0.260$\pm$0.017\\
Q1009+2956&NAL QSO&6&0.008& 0.014$\pm$0.002&0.028$\pm$0.007&0.049$\pm$0.013&0.174$\pm$0.046\\
Q1700+6416&NAL QSO&7&0.024&0.042$\pm$0.005&0.092$\pm$0.007&0.119$\pm$0.018&1.934$\pm$0.392\\
Q1946+7658&NAL QSO&7&0.014&0.020$\pm$0.003&0.051$\pm$0.014&0.094$\pm$0.020&0.674$\pm$0.220\\
\hline
\multicolumn{7}{@{}l@{}}{\hbox to 0pt{\parbox{180mm}{\footnotesize
Notes ---
\footnotemark[*] Number of observing epochs.\\
\footnotemark[$\dagger$] Standard deviation of magnitude of mini-BAL and NAL quasars.\\
\footnotemark[$\ddagger$] Mean quasar variability.\\
\footnotemark[$\S$] Maximum quasar variability.\\
\footnotemark[$\|$] Mean quasar variability gradient in the quasar rest-frame.\\
\footnotemark[$\#$] Maximum quasar variability gradient in the quasar rest-frame.\\
\footnotemark[**] Cannot be calculated because our sample was limited to two epochs.\\
\footnotemark[$\dagger \dagger$] Confidence level of quasar variability is below than 
3$\sigma$.\\
}\hss}}
\end{tabular}}
\end{center}
\end{table*}

\clearpage

\begin{table*}
\begin{center}
 \tbl{Observed frame equivalent width of C$\emissiontype{IV}$ mini-BAL
   in the HS1603+3820 spectrum}{%
\begin{tabular}{ccccc}
\hline
Observing Epoch$^{*}$&
$v_{shift}$&
$\Delta t_{\rm rest}^{\dagger}$&
${\rm EW}_{\rm C \emissiontype{IV}} ^{\ddagger}$&
 Detection Significance \\
 &(km s$^{-1}$)& & (\AA)& \\
\hline
 1 &$\sim 9,500$&0&11.3$\pm$3.1&3.6$\sigma$\\
 2 &&180.2&14.2$\pm$1.8&7.8$\sigma$\\
 3 &&258.6&17.3$\pm$2.8&6.2$\sigma$\\
 4 &&284.0&13.2$\pm$2.3&5.6$\sigma$\\
\hline
\multicolumn{5}{@{}l@{}}{\hbox to 0pt{\parbox{85mm}{\footnotesize
Notes ---
\footnotemark[*] Defined as in Table~3. 
\footnotemark[$\dagger$] Time delay from the first observation in the absorber 
rest-frame. Zero denotes the first observation epoch. 
\footnotemark[$\ddagger$] Equivalent width of C$\emissiontype{IV}$ mini-BAL in the 
observed frame.\\
}\hss}}
\end{tabular}}
\end{center}
\end{table*}

{\setlength\doublerulesep{5pt}  
\begin{table*}
\footnotesize
\begin{center}
\tbl{Power-law and asymptotic fitting parameters of 
structure functions}{%
\label{table:tab7}
\begin{tabular}{cccccc}
\hline
\multicolumn{2}{c}{}&
\multicolumn{2}{c}{$S_p$}&
\multicolumn{2}{c}{$S_a$}\\
\cmidrule(lr){3-4}
\cmidrule(lr){5-6}
Quasars&
Authors&
$\gamma$&
$S(\Delta \tau =$100d)&
$\Delta \tau_a$ (Asymptotic)&
$V_a$
\\
&
&
&
(mag)&
(day)&
(mag)
\\
\hline
\multicolumn{6}{c}{SDSS $u$-band}\\
\hline
mini-BAL quasars &this work&0.785$\pm$0.109&0.129$\pm$0.037&---$^{\dagger}$&---$^{\dagger}$\\
NAL quasars &this work&0.422$\pm$0.345&---$^{*}$&12.282$\pm$10.090&0.139$\pm$0.026\\
All of our quasars&this work&0.410$\pm$0.115&0.135$\pm$0.076&49.362$\pm$15.210&0.169$\pm$0.019\\
SDSS 7886 quasars&W08&0.435&0.173$\pm$0.001&---&---\\
\hline
\multicolumn{6}{c}{SDSS $g$-band}\\
\hline
mini-BAL quasars &this work&0.426$\pm$0.078&0.078$\pm$0.036&37.980$\pm$15.640&0.090$\pm$0.016\\
NAL quasars &this work&0.210$\pm$0.071&0.078$\pm$0.067&13.537$\pm$6.981&0.076$\pm$0.008\\
All of our quasars&this  work&0.264$\pm$0.056&0.080$\pm$0.043&20.768$\pm$7.478&0.082$\pm$0.008\\
SDSS 25,710 sample &VB04& 0.293$\pm$ 0.030 &---&51.9$\pm6.0^{\ddagger}$&0.168$\pm$0.005\\
SDSS 7886 quasars &W08&0.479&0.147$\pm$0.001&---&---\\
\hline
\multicolumn{6}{c}{SDSS $i$-band}\\
\hline
mini-BAL quasars &this work&0.446$\pm$0.263&---$^{*}$&18.870$\pm$9.088&0.073$\pm$0.008\\
NAL quasars &this work&0.432$\pm$0.111&---$^{*}$&---$^{*}$&---$^{*}$\\
All of our quasars&this work&0.432$\pm$0.121&---$^{*}$&---$^{\dagger}$&---$^{\dagger}$\\
SDSS 25,710 sample &VB04&0.303$\pm$0.035&---&62.6$\pm8.3^{\ddagger}$&0.139$\pm$0.005\\
SDSS 7886 quasars &W08&0.436&0.108$\pm$0.001&---&---\\
\hline
\multicolumn{6}{@{}l@{}}{\hbox to 0pt{\parbox{180mm}{\footnotesize
Notes --- 
\footnotemark[*] Unphysical values were obtained. \\
\footnotemark[$\dagger$] The data cannot be properly fitted by an asymptotic function.\\
\footnotemark[$\ddagger$] Data in VB04 not explicitly given to two decimal places. \\
}\hss}}
\end{tabular}}
\end{center}
\end{table*}
}

\clearpage

\begin{table*}
\begin{center}
 \tbl{Distribution properties of color variability versus quasar variability}{%
 \label{table:tab4}
 \begin{tabular}{cccc}
\hline
Distribution&$N^{*}$&$r^{\dagger}$&$a^{\ddagger}$ \\
\hline
\multicolumn{4}{c}{mini-BAL Quasar}\\
\hline
$\Delta (u-g)$-$\Delta u$&21&0.821&0.527$\pm$0.064\\
$\Delta (u-i)$-$\Delta u$&14&0.781&1.034$\pm$0.121\\
$\Delta (g-i)$-$\Delta g$&26&0.570&0.674$\pm$0.048\\
\hline
\multicolumn{4}{c}{NAL Quasar}\\
\hline
$\Delta(u-g)$-$\Delta u$&28&0.891&0.601$\pm$0.041\\
$\Delta(u-i)$-$\Delta u$&22&0.962&0.741$\pm$0.042\\
$\Delta(g-i)$-$\Delta g$&64&0.882&0.830$\pm$0.038\\
\hline
\multicolumn{4}{@{}l@{}}{\hbox to 0pt{\parbox{85mm}{\footnotesize
Notes ---
\footnotemark[*] Number of data points.\\
\footnotemark[$\dagger$] Pearson product-moment correlation coefficient.\\
\footnotemark[$\ddagger$] Slope of regression line.
}\hss}}
\end{tabular}}
\end{center}
\end{table*}

\begin{table*}
\begin{center}
 \tbl{Color variability properties of mini-BAL and NAL quasars}{%
\label{table:tab5}
 \begin{tabular}{cccccccc}
\hline
Color&
$\sigma_{\Delta c}^{*}$&
 $\langle \Delta C \rangle^{\dagger}$ &
 $|\Delta C_{\rm max}|^{\ddagger}$&
Quasar$^{\S}$&
 $\langle \Delta C/\Delta t_{\rm rest} \rangle^{\|}$ &
 $(|\Delta C/\Delta t_{\rm rest}|)_{\rm  max}^{\#}$ &
Quasar$^{**}$\\
&
(mag)&
(mag)&
(mag)&
&
(mag/yr)&
(mag/yr)\\
\hline
\multicolumn{8}{c}{mini-BAL Quasar}\\
\hline
$\Delta(u-g)$&0.057&0.058$\pm$0.010&0.184$\pm$0.051&Q1157+014&0.161$\pm$0.040&0.718$\pm$0.200&UM675\\
$\Delta(u-i)$&0.069&0.092$\pm$0.016&0.241$\pm$0.046&Q1157+014&0.305$\pm$0.048&0.482$\pm$0.104&HS1603+3820\\
$\Delta(g-i)$&0.038&0.051$\pm$0.007&0.136$\pm$0.023&Q1157+014&0.174$\pm$0.028&3.952$\pm$0.621&Q1157+014\\
\hline
\multicolumn{8}{c}{NAL Quasar}\\
\hline
$\Delta(u-g)$&0.047&0.071$\pm$0.009&0.182$\pm$0.025&Q1700+6416&0.170$\pm$0.034&1.956$\pm$0.609&Q1700+6416\\
$\Delta(u-i)$&0.060&0.080$\pm$0.013&0.218$\pm$0.022&Q1700+6416&0.188$\pm$0.046&1.374$\pm$0.129&Q1700+6416\\
$\Delta(g-i)$&0.049&0.048$\pm$0.006&0.208$\pm$0.022&Q1946+7658&0.107$\pm$0.025&5.329$\pm$0.584&Q1946+7658\\
\hline
\multicolumn{6}{@{}l@{}}{\hbox to 0pt{\parbox{180mm}{\footnotesize
Notes ---
\footnotemark[*] Standard deviation of color amplitude.\\
\footnotemark[$\dagger$] Mean amplitude of color variability.\\
\footnotemark[$\ddagger$] Maximum amplitude of color variability.\\
\footnotemark[$\S$] Quasar with maximum color variability amplitude.\\ 
\footnotemark[$\|$] Mean color variability gradient (per year).\\
\footnotemark[$\#$] Maximum color variability gradient (per year). \\
\footnotemark[**] Quasar with maximum color variability gradient. \\
}\hss}}
\end{tabular}}
\end{center}
\end{table*}


\end{document}